\documentclass[twocolumn]{aa}
\usepackage{graphicx}
\usepackage{txfonts}
\usepackage{natbib}
\usepackage{upgreek}
\usepackage{url}
\bibpunct{(}{)}{;}{a}{}{,}

\newcommand{\HII}{H\,{\sc ii}}
\newcommand{\kms}{~km\,s$^{-1}$}
\newcommand{\cm}{{\bf S8\,}}
\newcommand{\cn}{{\bf S9\,}}
\newcommand{\ca}{{\bf Sa\,}}
\newcommand{\cc}{{\bf Sc\,}}
\newcommand{\ce}{{\bf Se\,}}
\newcommand{\cf}{{\bf Sf\,}}
\newcommand{\cg}{{\bf Sg\,}}

\begin{document}

\title{
Massive envelopes and filaments in the NGC~3603 star forming region
  \thanks{Based in part on observations collected at the European
  Southern Observatory, Chile (Prop. No. 088.C-0093 and 090.C-0644).}
  \thanks{Observations were obtained with the Australia Telescope which 
         is funded by the Commonwealth of Australia for operations as a 
         National Facility managed by CSIRO.}
  }

\author{
  C.A.~Hummel\inst{1}      \and 
  T.~Stanke\inst{1}        \and  
  R.~Galv\'an-Madrid\inst{1,2}\and
  B.~S. Koribalski\inst{3}
}

\institute{
  European Southern Observatory,
  Karl-Schwarzschild-Str.~2, 85748 Garching bei M\"unchen, Germany 
  \thanks{Correspondence: {\tt chummel@eso.org}}
\and
  Centro de Radioastronom\'ia y Astrof\'isica, 
  Universidad Nacional Aut\'onoma de M\'exico, Morelia 58090, Mexico
\and
  CSIRO Astronomy and Space Science,
  Australia Telescope National Facility,
  P.O. Box 76, Epping, NSW 1710, Australia
  }

\date{Received: $<$date$>$; accepted: $<$date$>$; \LaTeX ed: \today}  

\abstract{
The formation of massive stars and their arrival on the zero-age
main-sequence occurs hidden behind dense clouds of gas and dust.  In the
giant \HII\ region NGC 3603, the radiation of a young cluster of OB stars
has dispersed dust and gas in its vicinity.  At a projected distance
of $2.5$ pc from the cluster, a bright mid-infrared (mid-IR) source
(IRS 9A) had been identified as a massive young stellar object (MYSO),
located on the side of a molecular clump (MM2) of gas facing the cluster.
We investigated the physical conditions in MM2,  based on APEX sub-mm
observations using the SABOCA and SHFI instruments, and archival ATCA
3 mm continuum and CS spectral line data. We resolved MM2 into several
compact cores, one of them closely associated with IRS 9A.  These are
likely infrared dark clouds as they do not show the typical hot-core
emission lines and are mostly opaque against the mid-IR background. The
compact cores have masses of up to several hundred times the solar mass
and gas temperatures of about 50 K, without evidence of internal ionizing
sources. We speculate that IRS 9A is younger than the cluster stars,
but is in an evolutionary state after that of the  compact cores.
}


\titlerunning{Sub-mm spectroscopy of NGC\,3603 IRS\,9A}

\authorrunning{C.A..~Hummel et al.}

\maketitle

\section{Introduction} \label{introduction}

The formation of high-mass stars is rapid \citep{zinnecker2007}
and leaves the new-born star still enshrouded in gas and dust.
Unlike the formation of low-mass stars via simple accretion disks,
theoretical studies show that filaments and nonaxisymmetric disks
\citep{krumholz2009} that funnel the radiative flux into the polar
directions \citep{2015ApJ...800...86K} provide protection for the
accreting gas against the uniquely intense stellar radiation pressure
until the final mass has been reached.

Looking for candidate massive young stellar objects (MYSO)
suitable for observation due to low foreground extinction,
\citet{2003A&A...404..255N} identified a bright infrared source
\citep[IRS 9A:][]{1977ApJ...213..723F}, on the side of a molecular clump
\citep[MM2:][]{2002A&A...394..253N} facing a massive cluster of OB stars
at the center of the \HII\ region NGC 3603.  \citet{2003A&A...404..255N}
estimated the mass of IRS 9A to be around 40 $M_\odot$, adopting a visual
extinction of 22 magnitudes due to circumstellar material gravitationally
bound to the MYSO.

\citet{2010A&A...520A..78V} combined 8-12 $\mu$m mid-infrared
(MIR) long-baseline (28-62 m) interferometry and single-dish (8 m)
sparse aperture synthesis to show that although the overall source size
is roughly 0.3", a line of sight towards a compact component of 0.06",
presumably at the center, must exist. This component could be associated
with the warm inner regions of an accretion disk, being photo-evaporated
by a newly formed O star if one considers the emission lines seen in a
MIR Spitzer spectrum of IRS 9A  \citep{2008ApJ...680..398L}.
Adopting the distance to the cluster, 7.2 kpc \citep{1989A&A...213...89M}
for the distance to MM2 and IRS 9A \citep[see also the discussion
in][]{1999AJ....117.2902D}, 1" corresponds to 7200 AU.

The (virial) mass of the molecular clump MM2 was determined by
\citet{2002A&A...394..253N} from observations of CS lines to be around
1500 solar masses. Due to the close association of the MYSO IRS 9A and
MM2 we performed observations at sub-mm radio wavelengths to
penetrate the obscuring dust and to determine total dust masses and
physical conditions in the emission area.

In this paper we present the resulting images of MM2 with a resolution
of a few arc-seconds obtained from interferometric observations of
the CS (2--1) line at mm-wavelength and from single-dish observations
of the sub-mm continuum emission.  The images resolve MM2 into seven
individual sources, one of which appears associated with IRS 9A.
We also obtained sub-mm spectra of the line emission at the position
of IRS 9A and a nearby source not detected in the mid-infrared
which allow us to draw conclusions on the total mass contained in
IRS 9A and the physical conditions inside it.  We discuss our
results in the context of sequential star formation in NGC 3603
\citep{1999AJ....117.2902D,2015ApJ...799..100D}.

\section{Observations and data reduction} \label{observations}

\subsection{ATCA $3$ mm interferometry}

We used archival data obtained with the Australia Telescope Compact
Array (ATCA) in its most compact configuration of molecular lines in
selected regions of the giant \HII\ region NGC~3603 that were carried
out in August 2005. This configuration has antennas on the northern spur
as well as the east-west track with baselines ranging from 31 to 89\,m.
For details of the observations see Table~\ref{atca}. Two transitions,
CS(2--1) and C$^{34}$S(2--1), had been observed simultaneously with a
bandwith of 16 MHz each, split into 256 channels, giving a channel width
of 0.19\kms.  The rest frequencies of the two CS lines are 97.980968
GHz and 96.412982 GHz, respectively. We mosaiced the molecular clump
MM2 \citep{2002A&A...394..253N} in the southern part of NGC~3603, using
a $3 \times 3$ grid, covering an area of $1\arcmin \times 1\arcmin$.
To achieve Nyquist sampling the pointings had been separated by 15\arcsec,
about half the primary beam width. The weather conditions (during the day)
had been reasonable, and a system temperatures of 220 K was measured
($\pm$10\%) for antennas 2--5, but a significantly higher value of
320 K for antenna 1.  The antenna efficiency is around 25\% at these
frequencies. The observing procedure had comprised a setup on a strong
3\,mm calibrator at high elevation such as PKS\,0537--441, bandpass
calibration on PKS\,1253--055, and phase calibration on PKS\,1045--62 and
$\eta$ Carina. The phase calibration was difficult as PKS\,1045--62 was
very weak and $\eta$ Carina is extended. Pointing had been performed on
$\eta$ Carina every hour, and the system temperature was obtained every
30 minutes with a paddle above the 3\,mm horn.

As part of a flux monitoring program for 3\,mm calibrators, strong sources
such as PKS\,1253--055 (3C\,279) and PKS\,0537--441 are observed regularly
with the ATCA. PKS\,1253--055 was measured to have a flux density of
14.8 Jy and 13.5 Jy (at 94.5 GHz) on the 8th and 17th of August, 2005,
respectively.  We used these values for the absolute flux calibration.

Data reduction was carried out in the software package {\sc miriad}
\citep{1995ASPC...77..433S} using standard procedures.

We used the program {\sc plboot} to determine gain calibration factors
of 2.64 and 2.45 for 97969 \& 96401 MHz, respectively, from observations
of Mars with the H214 array on the 19th of August, 2005. Using this factor,
we derived 13.5 Jy for PKS\,1253--055, in excellent agreement with its
3\,mm flux as determined on the 17th of August as part of the calibrator
monitoring project (see http://www.narrabri.atnf.csiro.au/calibrators/).

\begin{figure} 
\centering
\begin{tabular}{c}
\includegraphics[width=1.0\columnwidth]{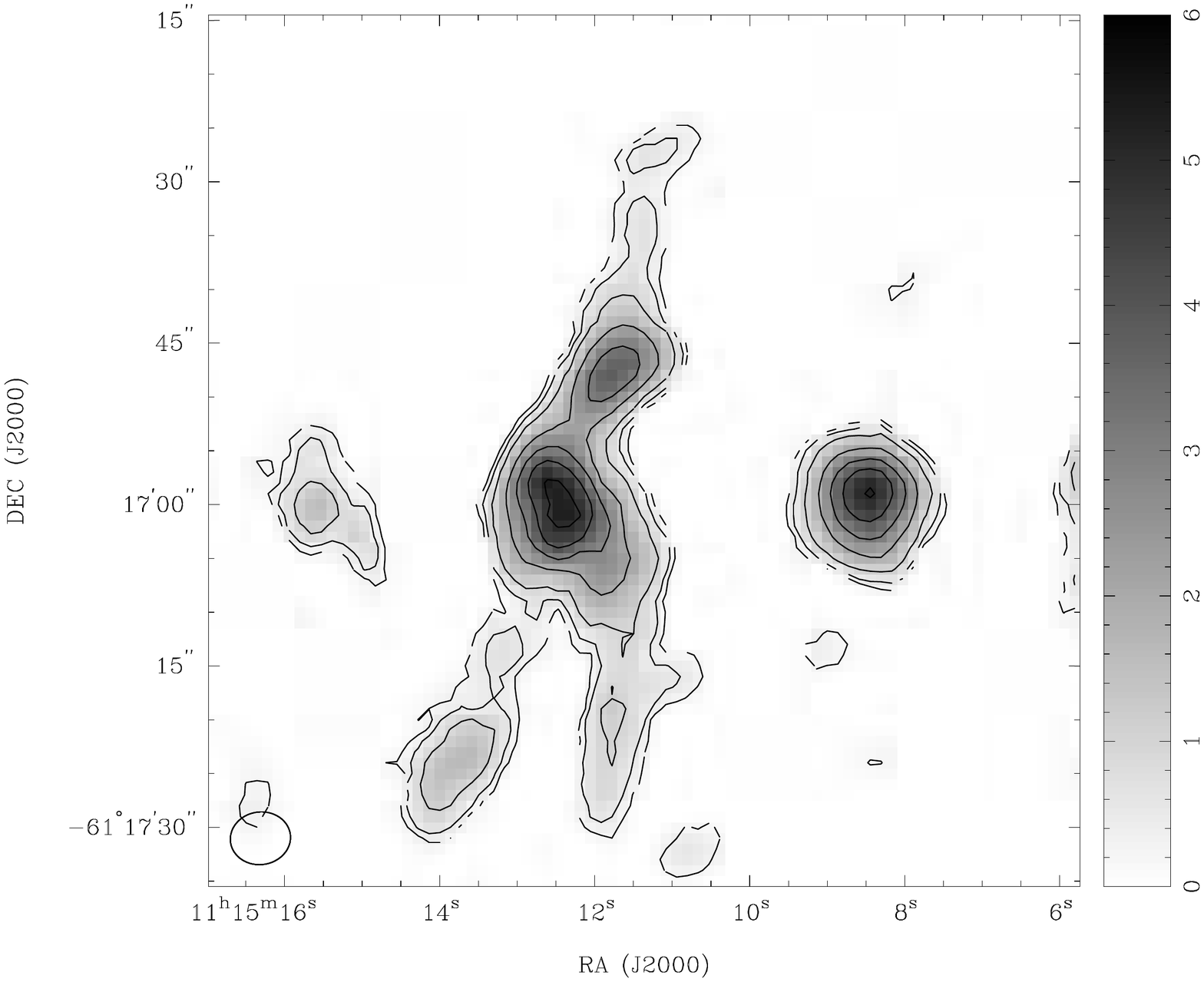} \\
\includegraphics[width=1.0\columnwidth,angle=0]{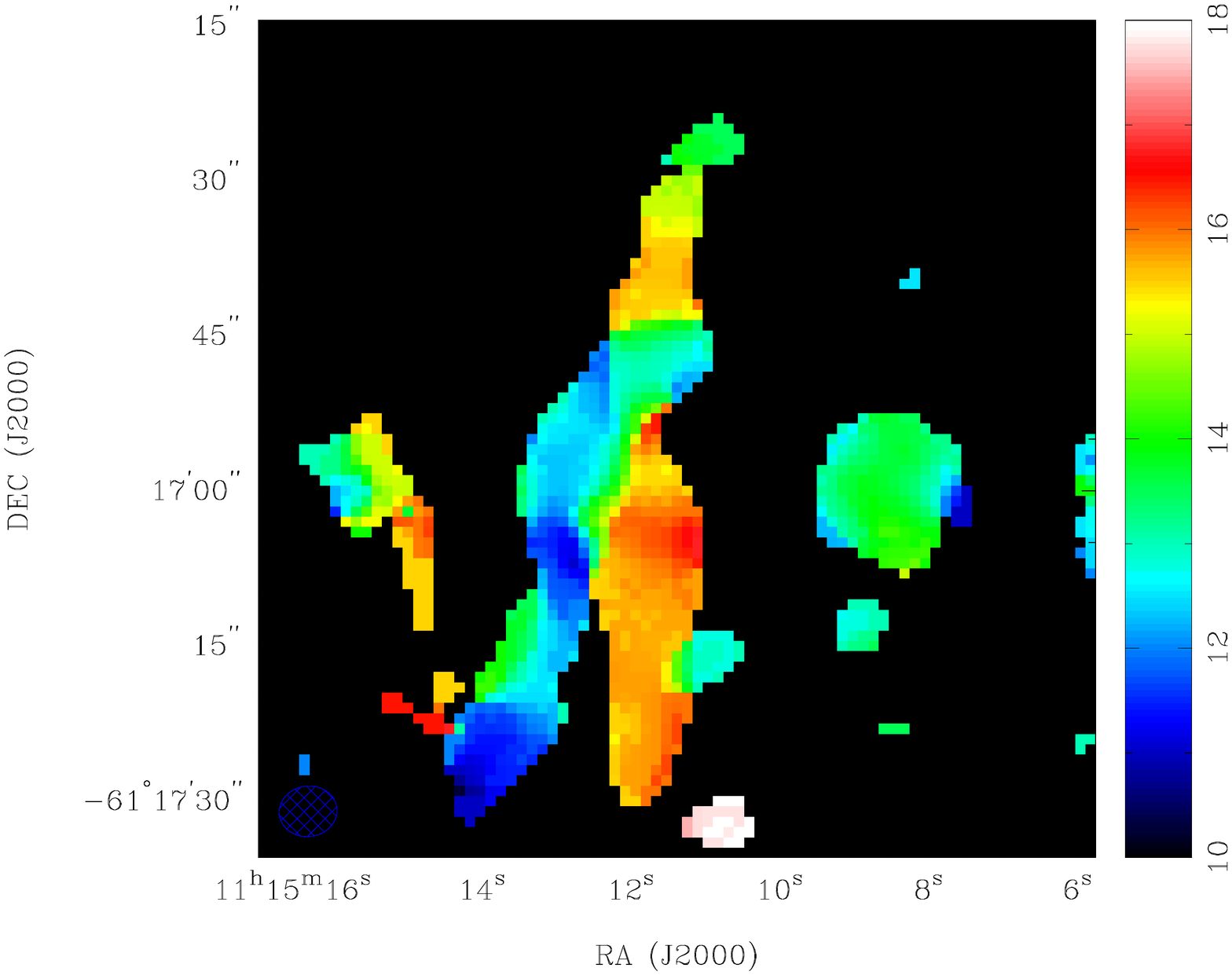} \\
\end{tabular}
\caption{ATCA CS(2--1) molecular line emission in NGC~3603 MM2. {\bf Top}:
integrated intensity map; the contour levels are 0.25, 0.5, 1,
2, 3, 4, and 5 Jy\,beam$^{-1}$\kms. {\bf Bottom}: mean velocity
field; the synthesized beam ($5.6" \times 4.9"$) is displayed at the
bottom left of each panel, and the grey scale/color wedges for data
values on the right of each panel.}
\label{atca_img_cs}
\end{figure}
\begin{figure} 
\centering
\begin{tabular}{c}
\includegraphics[width=1.0\columnwidth]{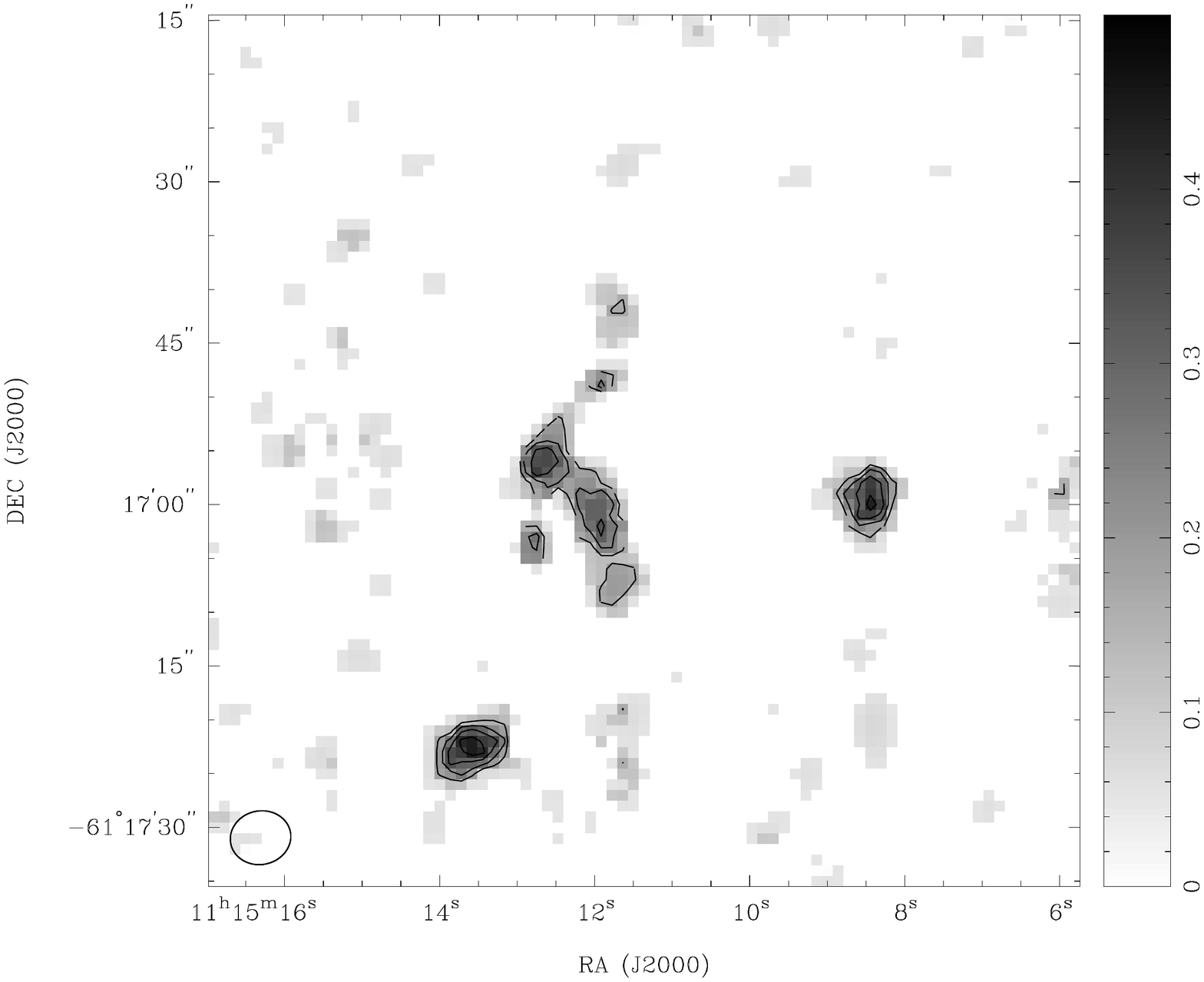} \\
\includegraphics[width=1.0\columnwidth]{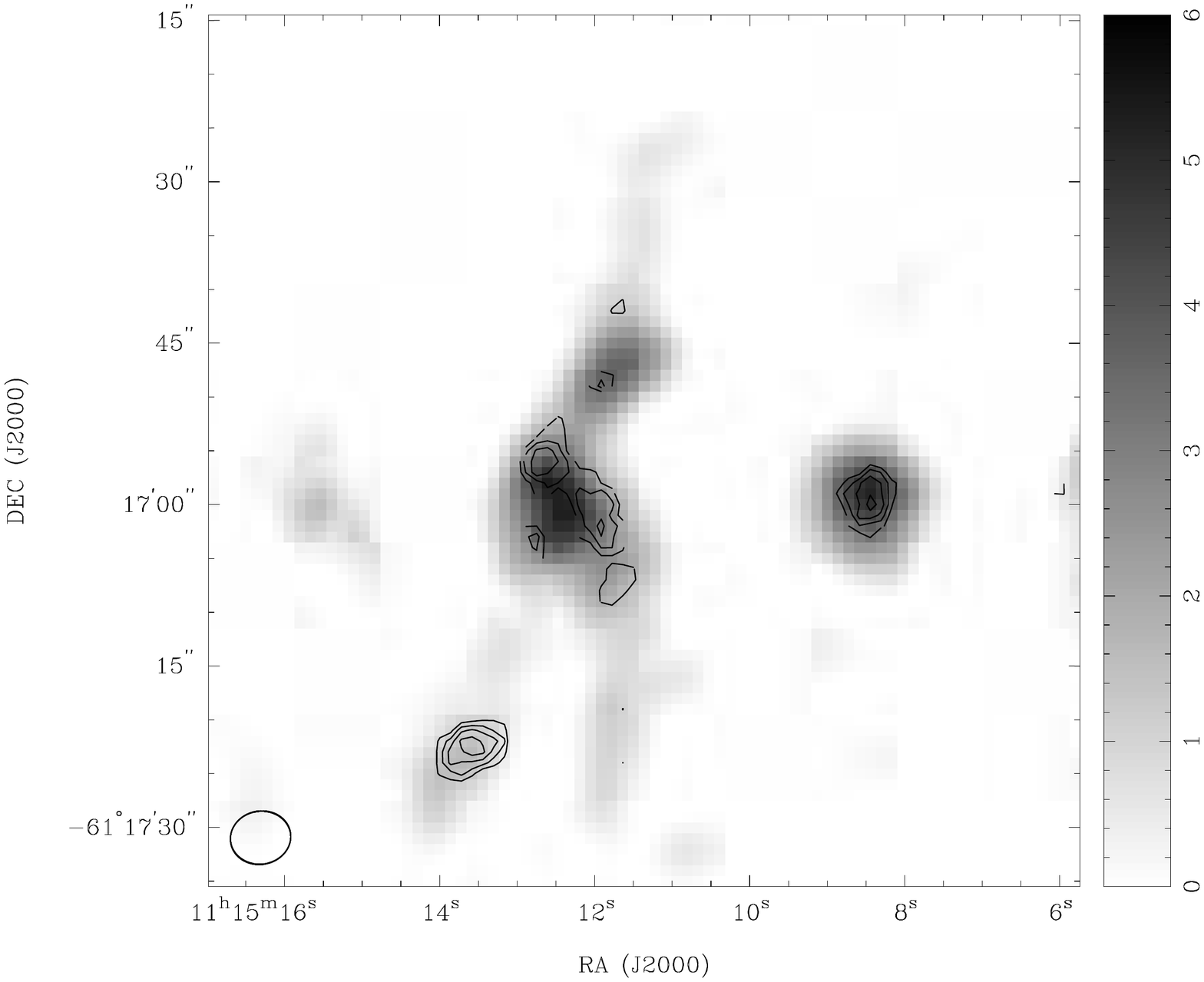} \\
\end{tabular}
\caption{ATCA C$^{34}$S(2--1) molecular line emission in NGC~3603 MM2.
{\bf Top}: integrated intensity map; the contour levels are 0.16, 0.24,
0.32, and 0.40 Jy\,beam$^{-1}$\kms.
{\bf Bottom}: C$^{34}$S(2--1) contours overlaid onto CS(2--1) emission
map.  The synthesized beam ($5.6" \times 4.9"$) is displayed at the
bottom left of each panel, and the grey-scale wedge for data values on
the right of each panel.
}
\label{atca_img_c34s}
\end{figure}
\begin{figure} 
\begin{tabular}{cc}
\includegraphics[width=0.35\columnwidth,angle=-90]{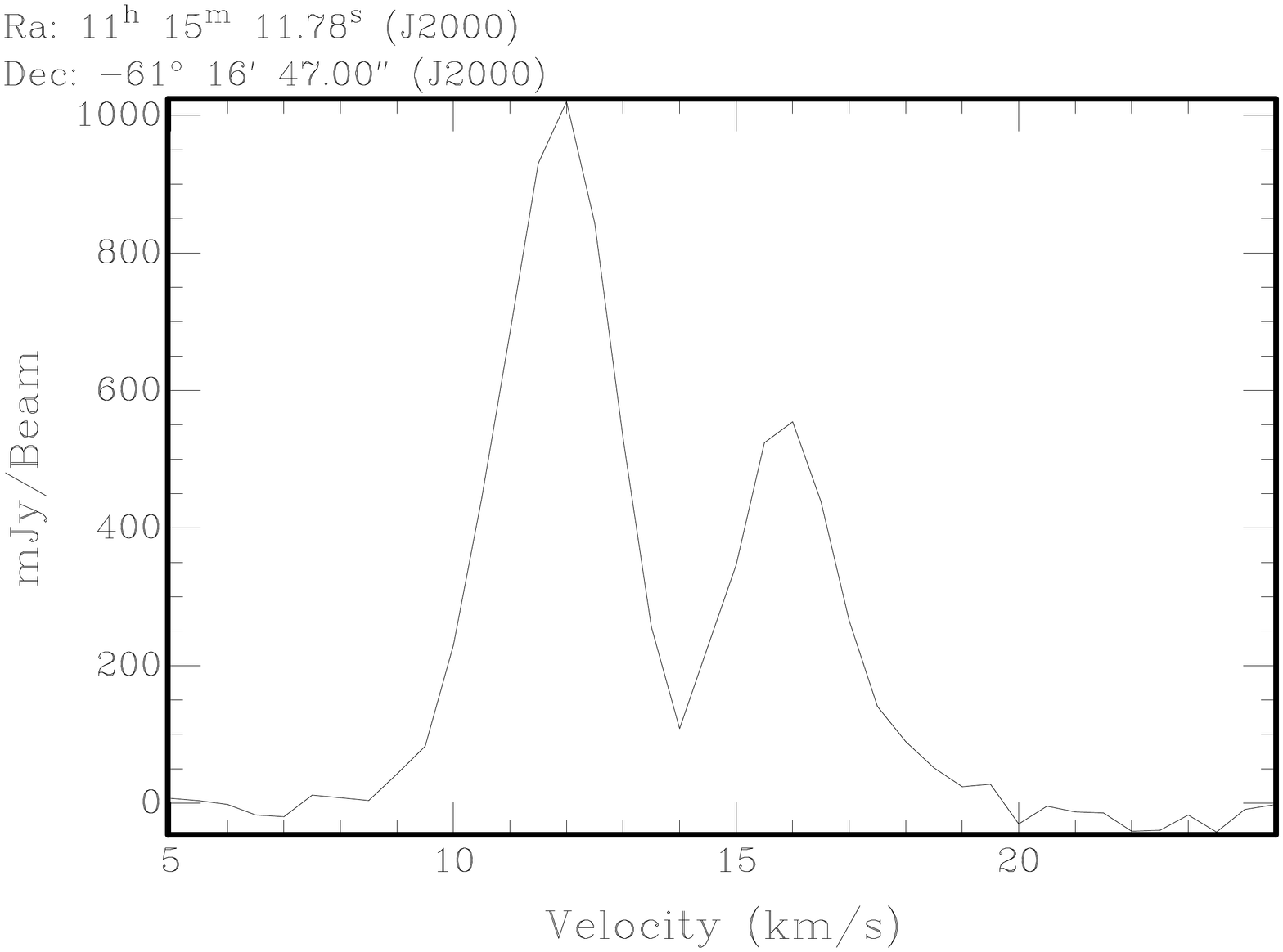}  &	
\includegraphics[width=0.35\columnwidth,angle=-90]{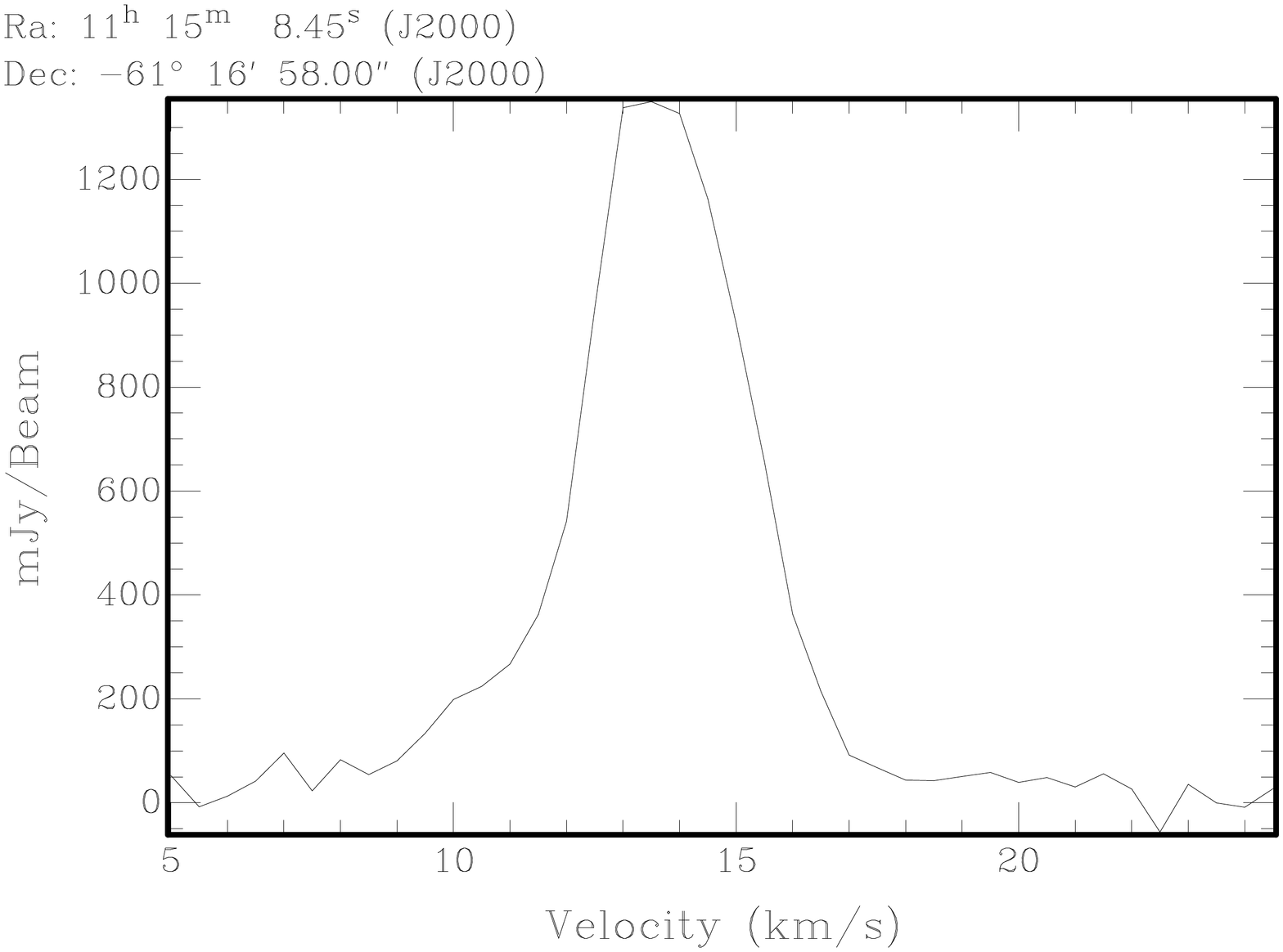}  \\	
\includegraphics[width=0.35\columnwidth,angle=-90]{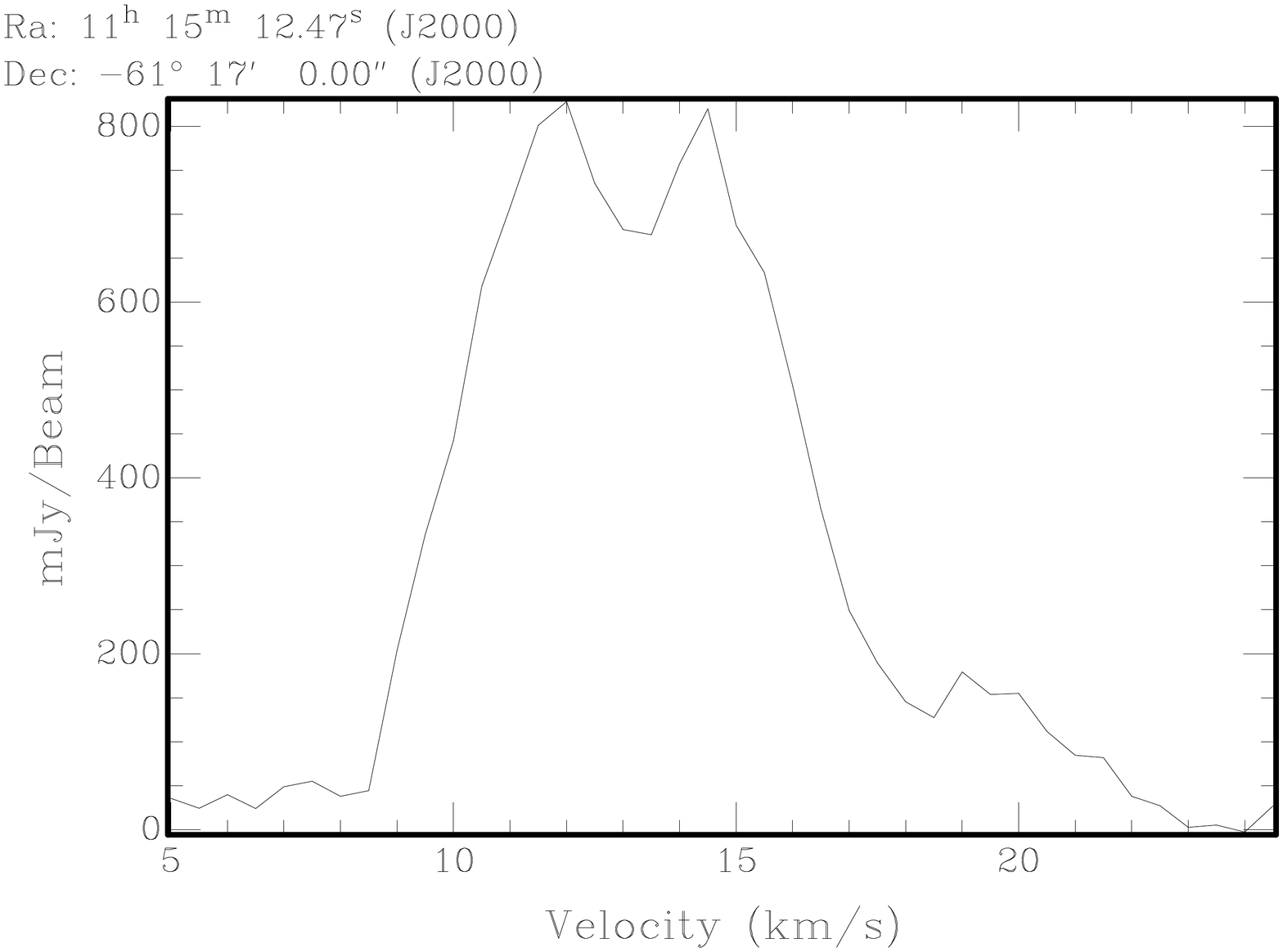} &	
\includegraphics[width=0.35\columnwidth,angle=-90]{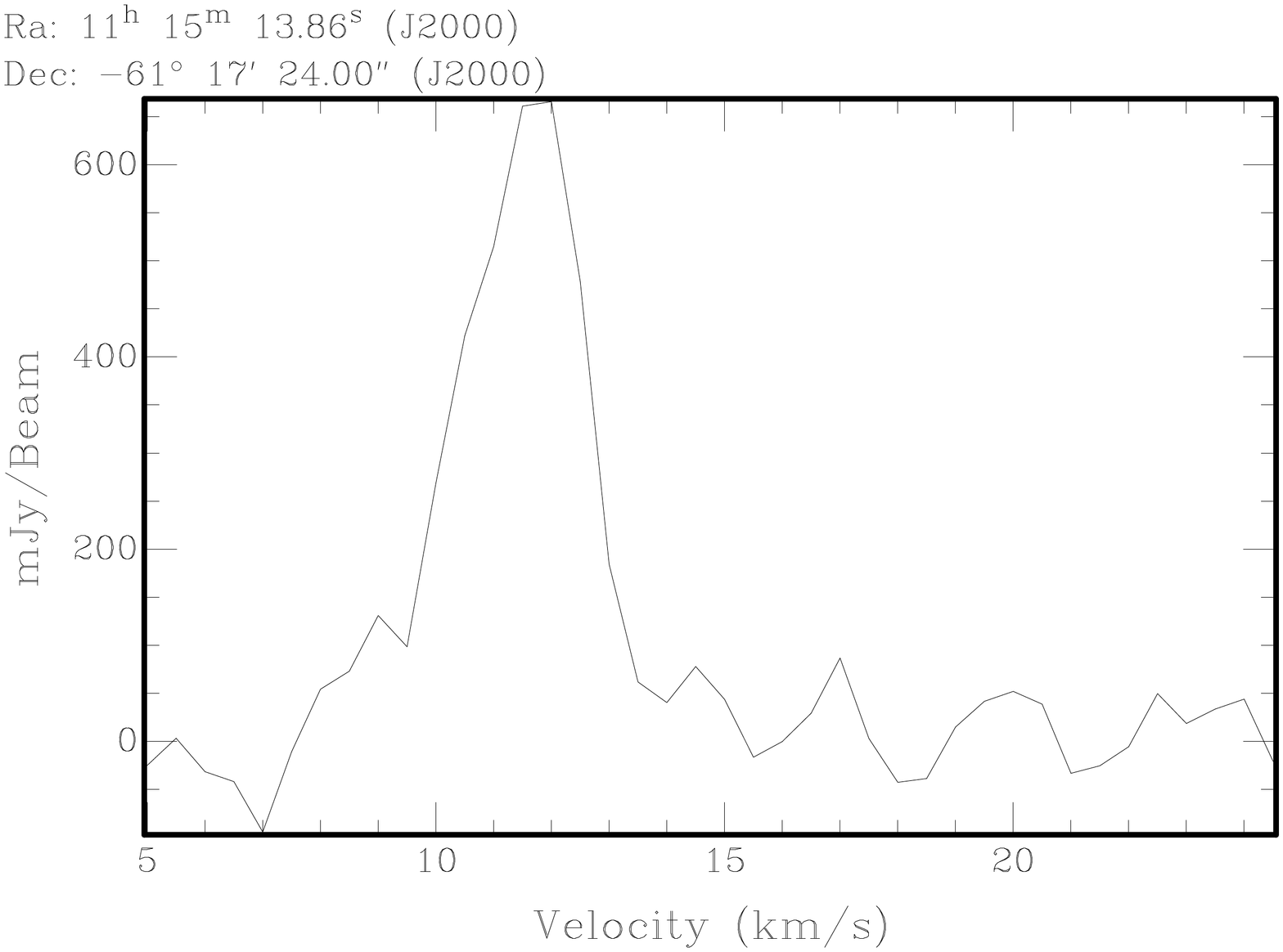}  \\	
\end{tabular}
\caption{ATCA CS(2--1) spectra at four selected positions (coordinates
given in the panels) within NGC~3603 MM2. From left to right and
top to bottom: sources \cn, \cc, \ca, \ce
of Fig.~\ref{saboca}.}
\label{atca_cs}
\end{figure}

Using natural weighting and mosaicing we made line cubes and continuum
maps for NGC~3603 MM2. We used a velocity resolution of 0.5\kms for
the cubes and derived the mean CS(2--1) velocity field.  No reliable
detections could be obtained in the continuum at 3\,mm above the
noise of $\sim$35 mJy/beam RMS measured in the line-free channels.
Figs.~\ref{atca_img_cs} and \ref{atca_img_c34s} show the CS (2--1) and
C$^{34}$S emission maps, respectively. Spectra of CS(2-1) emission for
selected positions are shown in Fig.~\ref{atca_cs}.

\begin{table}
\caption{ATCA 2005 observing parameters} 
\label{atca}
\begin{tabular}{lc}
\hline
Configuration   &    H75                \\
\hline
Obs. dates      & 2005, 5--7 August     \\
Antennas        & 5                     \\
Baselines       & 31,31,43,46,46,55,       \\
                & 77,77,82, and 89\,m   \\
Total obs. time & $2 \times 6$ h (MM2)  \\
$T_{\rm sys}$   & 200--350 K            \\ 
Bandwidth       & $2 \times 16$ MHz                 \\
No. of channels & $2 \times 256$                    \\
Channel width   & 0.19\kms                          \\
Velocity range  & --15 to +35\kms                   \\
Center freq. IF1& 97969 MHz \\
Center freq. IF2& 96401 MHz \\
Primary beam    & 29 arcsec                         \\
Field center    & $\alpha$(J2000) =
                  11:15:11.5                         \\
Field center    & $\delta$(J2000) =
                  --61:16:55                  \\
\hline
Synthesized beam& $5.6" \times 4.9"$    \\
RMS per channel & 35 mJy\,beam$^{-1}$  \\
Channel width   & 0.5\kms\ (after smoothing)\\
\hline
Flux \& bandpass&                        \\
~~~calibrator   &PKS\,1253--055 (14.8 Jy)\\
Phase calibrator&PKS\,1045--62 (0.45 Jy) \\
                & EtaCar (9.0 Jy)        \\
\hline
\end{tabular}
\end{table}

\subsection{SABOCA sub-mm imaging}

\begin{figure}
\centering
\begin{tabular}{l}
\includegraphics[width=1.0\columnwidth]{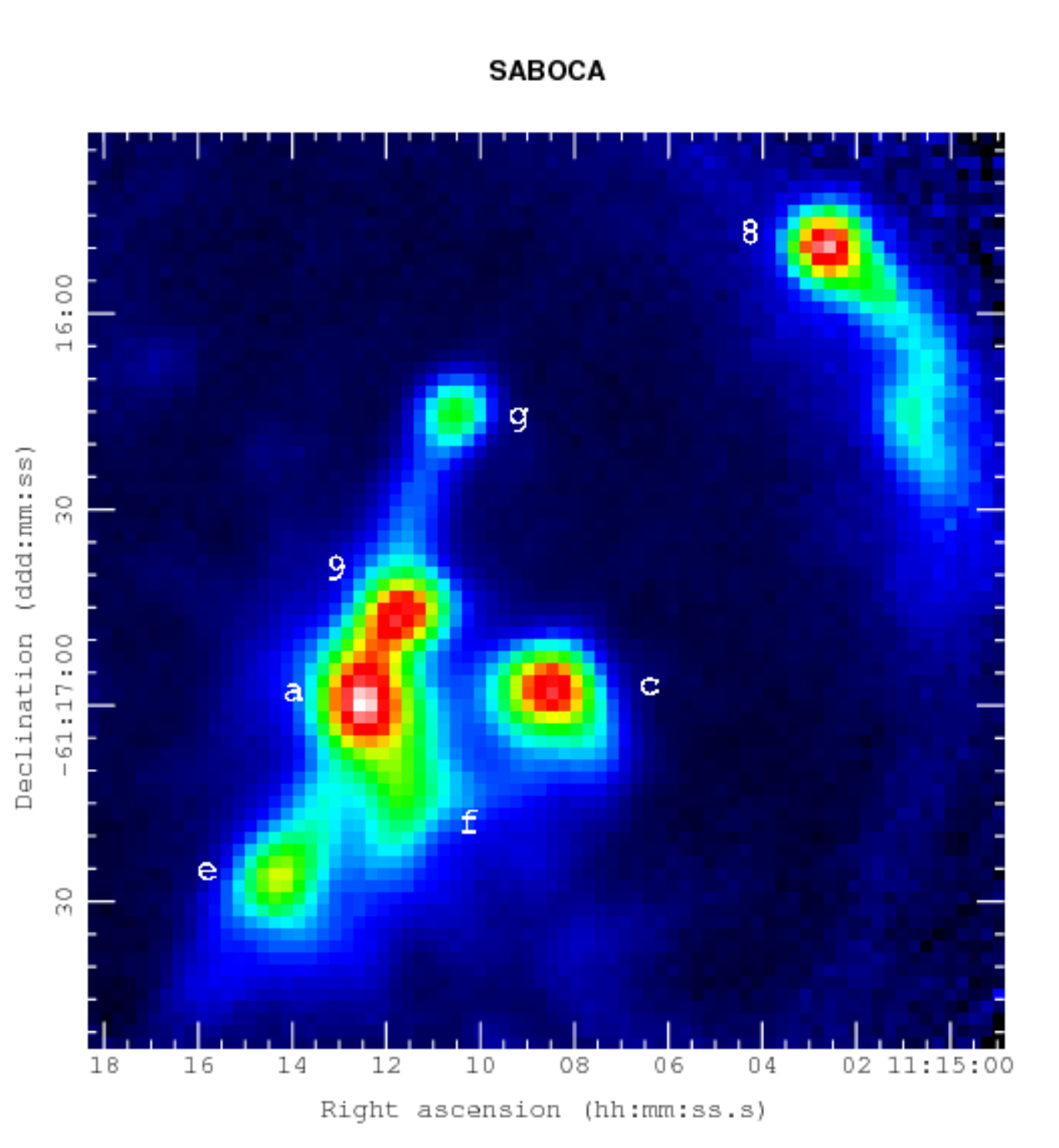}\\
\includegraphics[width=1.0\columnwidth]{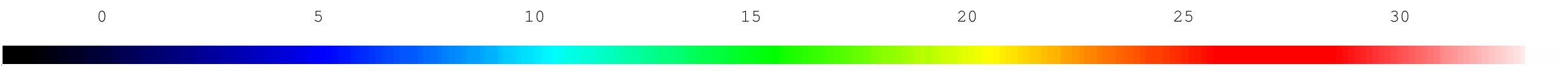}
\end{tabular}
\caption{SABOCA 350 $\mu$m image of NGC~3603 MM2. The source labeled
``{\bf 9}'' (\cn) is associated with IRS 9A.  This source
and the source labeled ``{\bf c}'' (\cc) were the targets for SHFI
spectroscopy. The source labeled ``{\bf g}'' (\cg) is at the tip of
the eastern pillar (see Figs.~1 of \citet{1999A&A...352L..69B} and
\citet{2002ApJ...571..366M}), while the source labeled ``{\bf 8}''
(\cm) is at the tip of the western pillar, both pointing towards the
OB cluster.  The wedge at the bottom shows the scale for the image flux
densities in Jy/Beam.
}
\label{saboca}
\end{figure}

Observations of NGC~3603 MM2 at 350 $\mu$m with the SABOCA bolometer
array \citep{2010Msngr.139...20S} attached to the APEX telescope on the
Chajnantor plateau in Chile were carried out on September 15, 2011,
providing a beam FWHM of 7.8". The precipitable water vapor column was
about 0.2 mm during the observations which lasted about 15 minutes for
a spiral raster map of about 1.5 arcminutes in radius centered on IRS
9A. For the calibration, a sky dip determined a zenith opacity of 0.736,
and absolute calibration was established with observations of VY CMa and
B13134 from the APEX primary and secondary calibrator list. The accuracy
of the absolute calibration is estimated to be 10\%. The map is shown
in Fig.~\ref{saboca}.

Seven distinct sources can be identified in the 350 $\mu$m SABOCA image
(Fig.~\ref{saboca}), some of which were already seen in the ATCA CS maps
(Figs.~\ref{atca_img_cs} and \ref{atca_img_c34s}).  The positions of the
sources and their flux densities were extracted by iteratively fitting
and subtracting two-dimensional Gaussian profiles, starting with the
brightest source. The total fluxes were derived from the fitted peak
flux and width of the Gaussian profiles, deconvolved from the beam width.
The results are summarized in Table~\ref{sources}.

\begin{table}
\caption{Properties of SABOCA sources. The sources labeled \cm
and \cn in the first column are the same as listed in Table~1 of
\citet{2008ApJ...680..398L}. Source \cn is associated with IRS
9A. The source size $\theta$ in column 5 is the geometric mean of
the axes of the deconvolved size. At the distance of IRS 9A, 10"
correspond to 0.33 pc. The data in the last three columns are described
in Section~\ref{massofthecompactcores}.
}
\label{sources}
\centering
\begin{tabular}{cccccccc}
\hline\hline
       & RA   &  Dec  & $F$ & $\theta$ & $\Delta$v& $M_F$     & $M_V$ \\
       &      &       & Jy  &  "       & km/s     & $M_\odot$ & $M_\odot$ \\
\hline
\ca    & 11:15:12.07 & -61:16:58.8 & 90 &  9.2        & 2.5 & 498 & 122 \\
\cm    & 11:15:02.88 & -61:15:51.5 & 55 &  6.7        & $-$ & 307 &   0 \\
\cc    & 11:15:08.41 & -61:16:57.2 & 67 &  8.2        & 3.2 & 369 & 180 \\
\cn    & 11:15:11.34 & -61:16:45.2 & 59 &  7.6        & 2.5 & 329 & 102 \\
\ce    & 11:15:13.73 & -61:17:24.3 & 46 &  8.5        & 2.5 & 258 & 114 \\
\cf    & 11:15:11.25 & -61:17:12.0 & 45 &  9.6        & $-$ & 249 &   0 \\
\cg    & 11:15:10.33 & -61:16:16.4 & 22 &  4.9        & $-$ & 122 &   0 \\
\hline
\end{tabular}
\end{table}

\subsection{SHFI sub-mm spectroscopy}

\begin{table}
\caption{SHFI observational parameters. The main beam efficiencies 
($\eta_{\rm mb}=0.75$ for APEX-1 and $0.73$ for APEX-2) and 
conversion factors ($39$ K/Jy for APEX-1 and $41$ K/Jy for APEX-2)
are taken from {\tt http://www.apex-telescope.org/telescope/efficiency/}}
\label{shfi}
\begin{tabular}{lcccccc}
\hline\hline
Receiver& Setting& Range 		& Noise 	& PWV \\
	&	&	GHz		& mK		& mm \\
\hline
APEX-1 	& 1	& 216.9 -- 220.9	& 10		& 2.0 \\
       	& 2	& 229.2 -- 233.2	& 30		& 2.0 \\
APEX-2 	& 3	& 346.1 -- 350.1	& 30		& 1.5 \\
       	& 4	& 353.2 -- 357.2	& 30		& 0.4 \\
       	& 5	& 345.8			& 50		& 0.3 \\
\hline
\end{tabular}
\end{table}

Observations with the SHFI instrument \citep{2008A&A...490.1157V}
attached to APEX were carried out in 2012. The observations with the
APEX Band-1 (211 - 275 GHz) receiver were carried out July 31, those with
the Band-2 receiver (275 - 370 GHz) August 1 (setting 3) and October 10
(setting 4). Finally, on October 12, a 25-pointing map was executed in
the CO (3-2) line (setting 5). Details are given in Table~\ref{shfi}.
For all observations, a velocity resolution of 0.5 \kms was chosen.

Each observation sequence (except for the map) included two back-to-back
pointings towards sources \cn and \cc (28" distance), sandwiched between two
observations of the sky at RA = 11:19:58.5 and DEC = -60:09:57.8 which
was selected for its low 100 $\mu$m emission (IRAS) a short distance away
(1.2 degrees). Even with the largest beam of 32" of APEX Band-1, the distance
between \cn and \cc of 28" is large enough to prevent mutual contamination.
However, sources \cn and \ca have a distance of 16" so that
the flux measured at position \cn might be contaminated by emission from
\ca, as half of the flux of \ca will be picked up by the 28" beam.

The spectra with a bandwidth of 4 GHz were reduced using the CLASS
software\footnote{\tt http://www.iram.fr/IRAMFR/GILDAS}. In order
to identify the lines, the systemic velocity of source {\bf c} was
measured from the optically thin isotopologues of CO (C$^{18}$O) to be
+13 \kms.  NIST recommended rest frequencies of the lines\footnote{\tt
http://physics.nist.gov/cgi-bin/micro/table5/start.pl} were used to
identify the lines listed in Table~\ref{brightlines}, together with their
properties and the antenna temperature peak ratios between sources \cc
and \cn ($T_{\rm c}/T_{\rm 9}$).

We used the WEEDS extension \citep{2011A&A...526A..47M} of CLASS to
confirm the line identifications on source \cn by checking for each
species the presence of lines in any of our wavelength settings. Using
the three Formaldehyde lines we detected in setting 1, we determined an
excitation temperature of 45 K as this value reproduced the measured line
ratios.  All lines were fit with this temperature, and a line width of
5 km/s.  Two of the lines, $^{13}$CO and C$^{18}$O, are extended by about
a factor of six over the other line emitting regions, which we adopted
at 5" in size (unresolved by APEX). Increasing the number density of
the species instead would have led to saturation of the lines.

Two lines near 230.232 GHz and 230.841 GHz remained unidentified. The
closest matches of lines from SO$^{17}$O, CH$_3$OCH$_3$, and CH$_2$CHCN,
respectively, predicted more lines of these species to be seen in setting
4, but were not detected.

\begin{figure}
\begin{center}
\begin{tabular}{cc}
\includegraphics[clip,width=0.5\columnwidth]{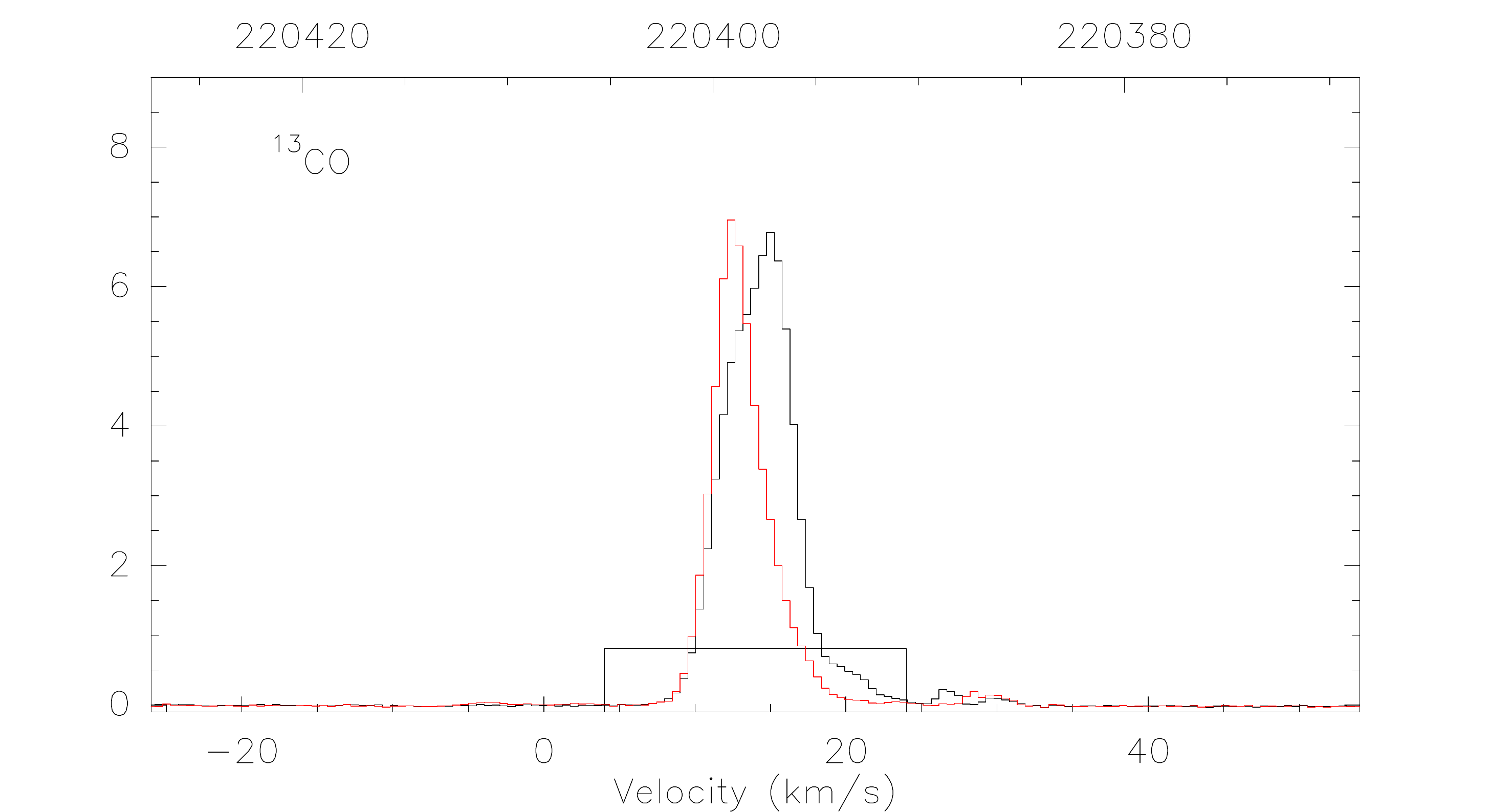} &
\includegraphics[clip,width=0.5\columnwidth]{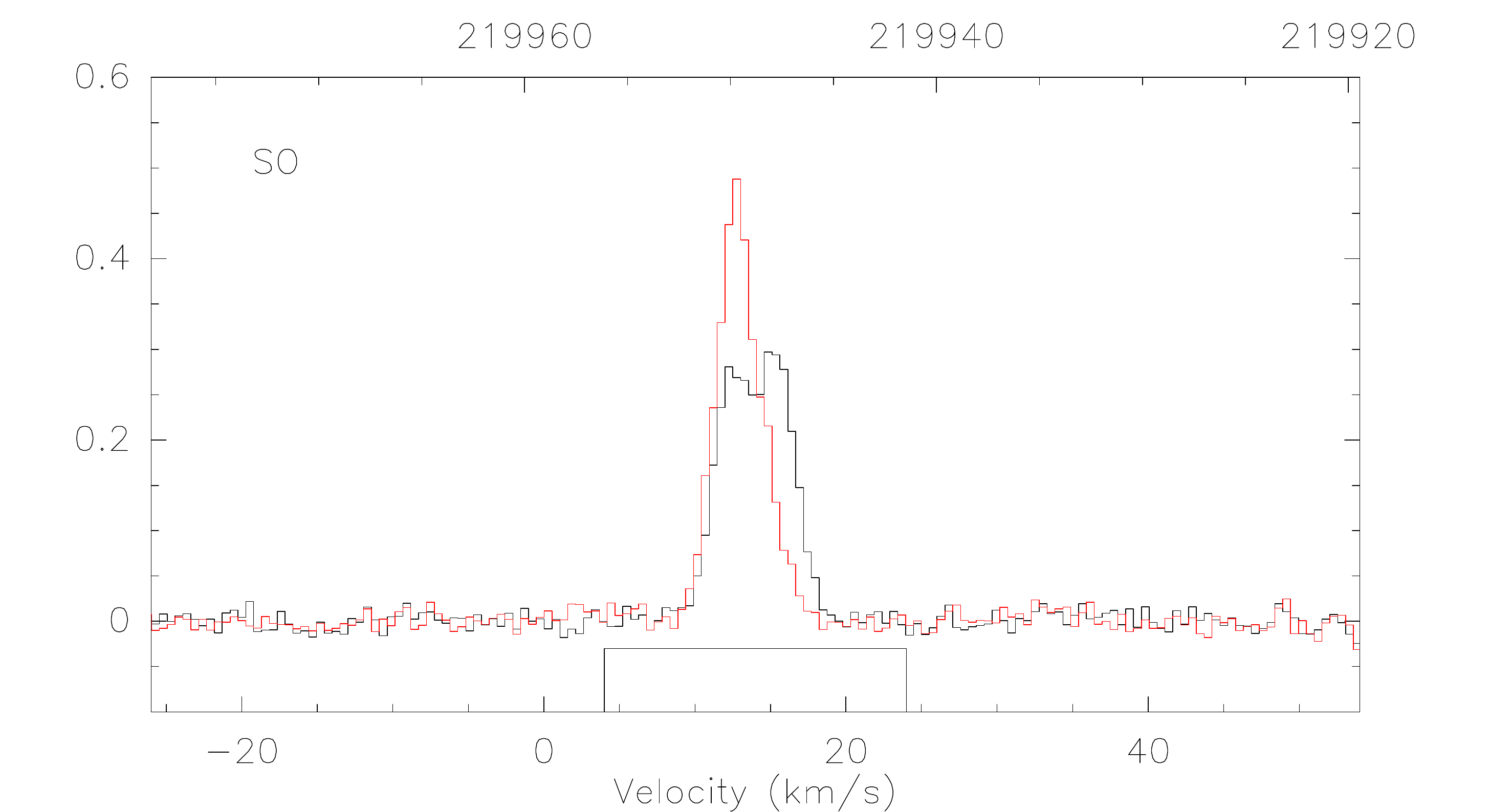} \\
\includegraphics[clip,width=0.5\columnwidth]{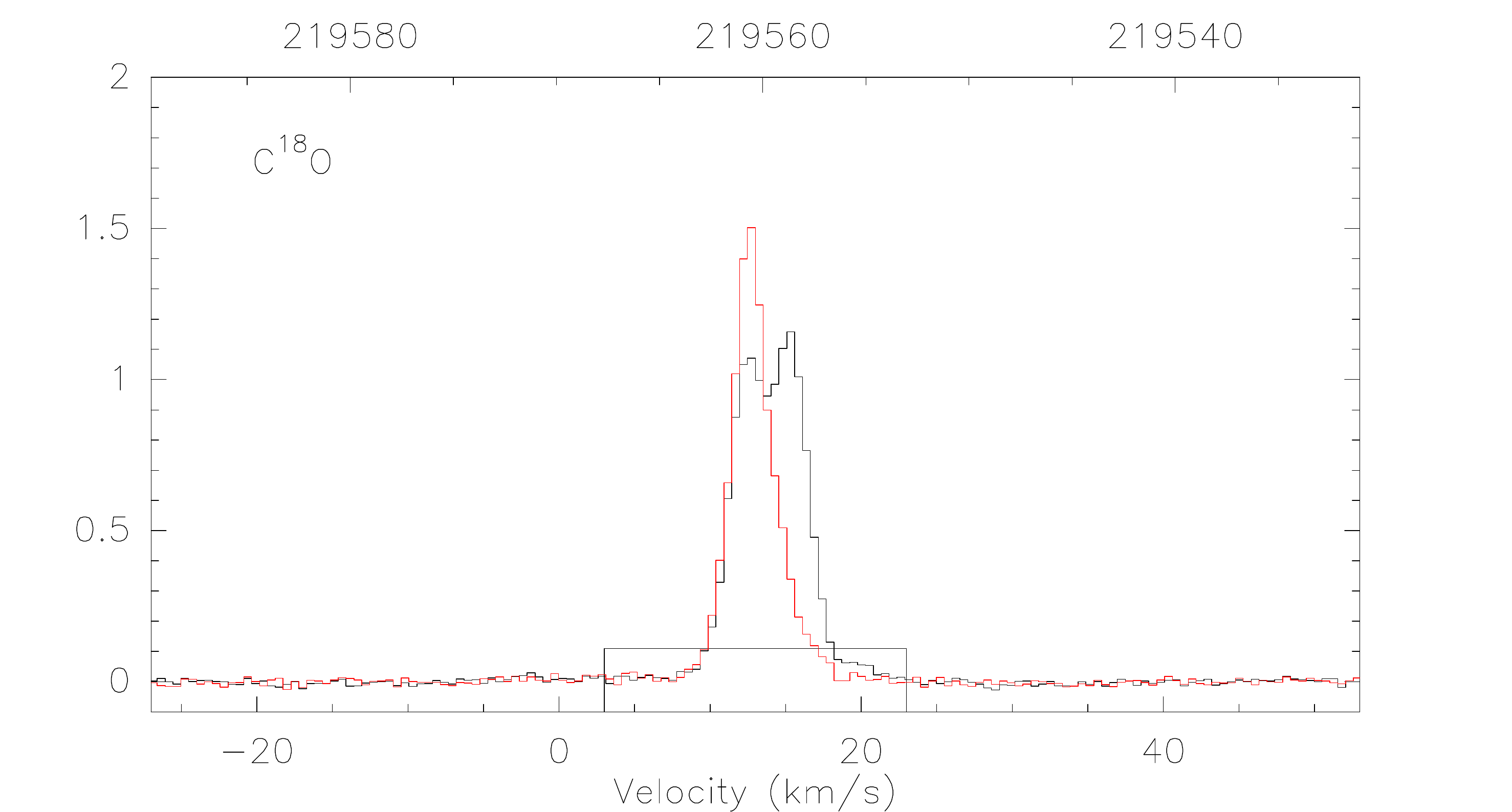} &
\includegraphics[clip,width=0.5\columnwidth]{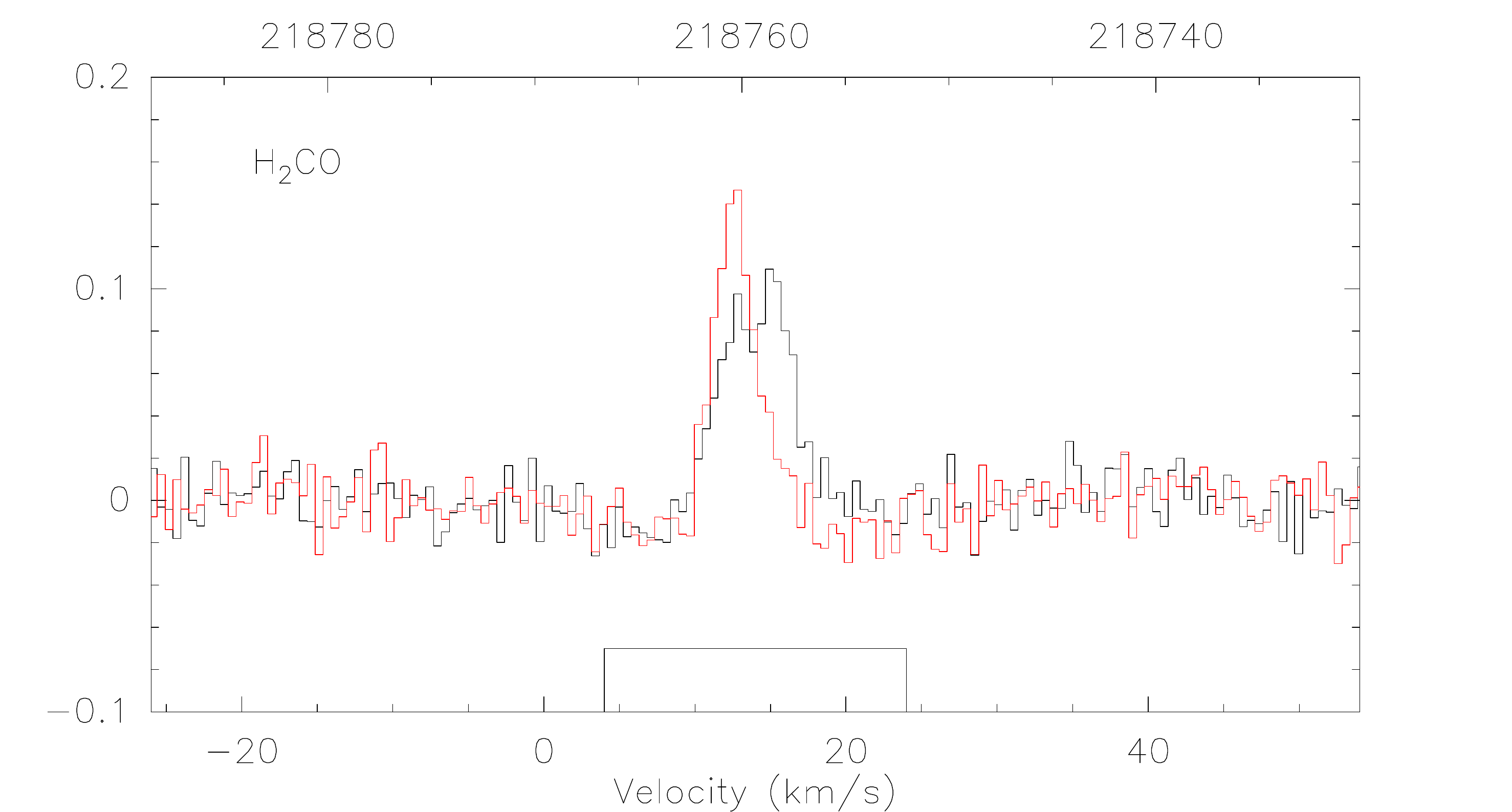} \\
\includegraphics[clip,width=0.5\columnwidth]{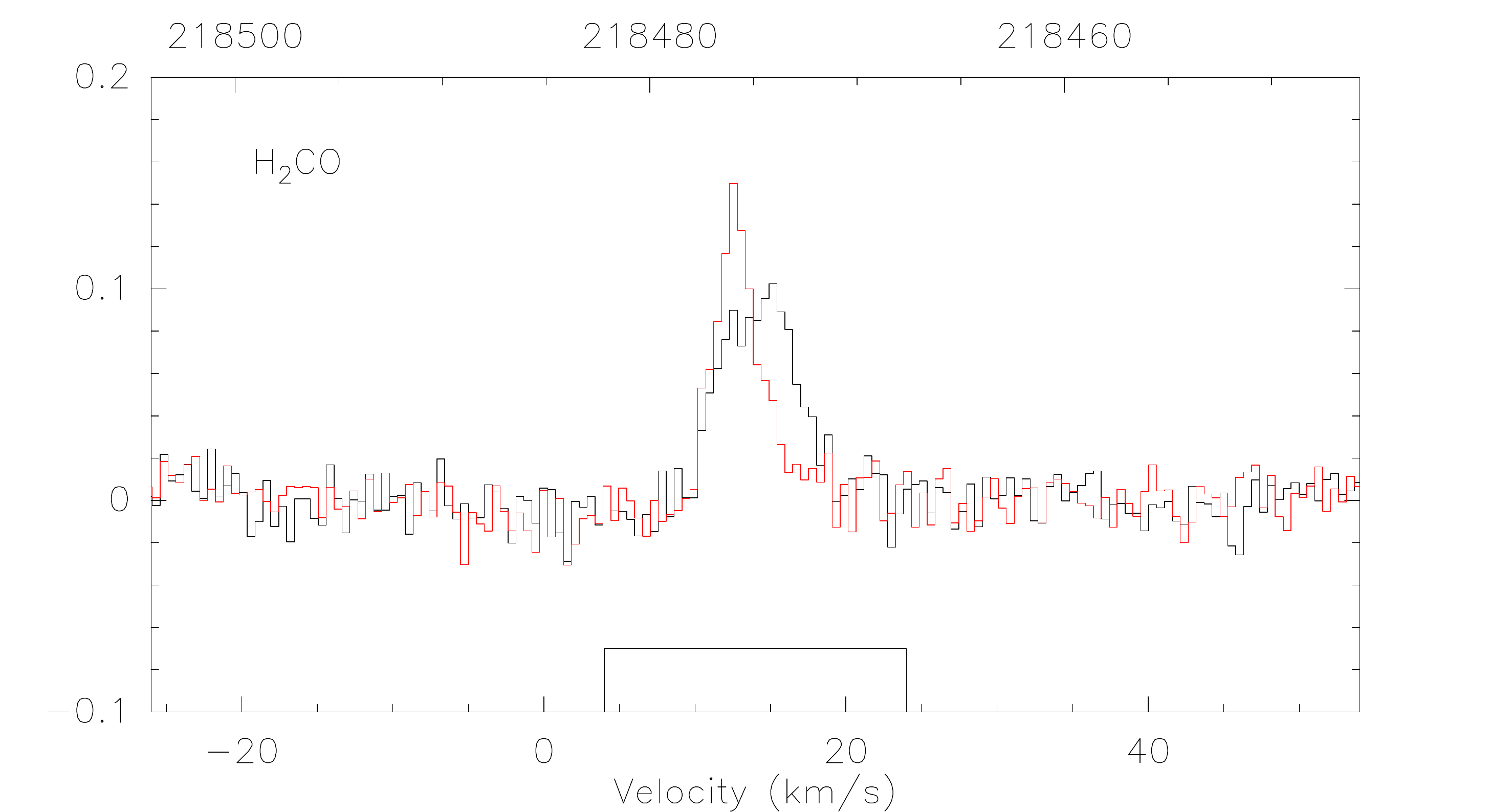} &
\includegraphics[clip,width=0.5\columnwidth]{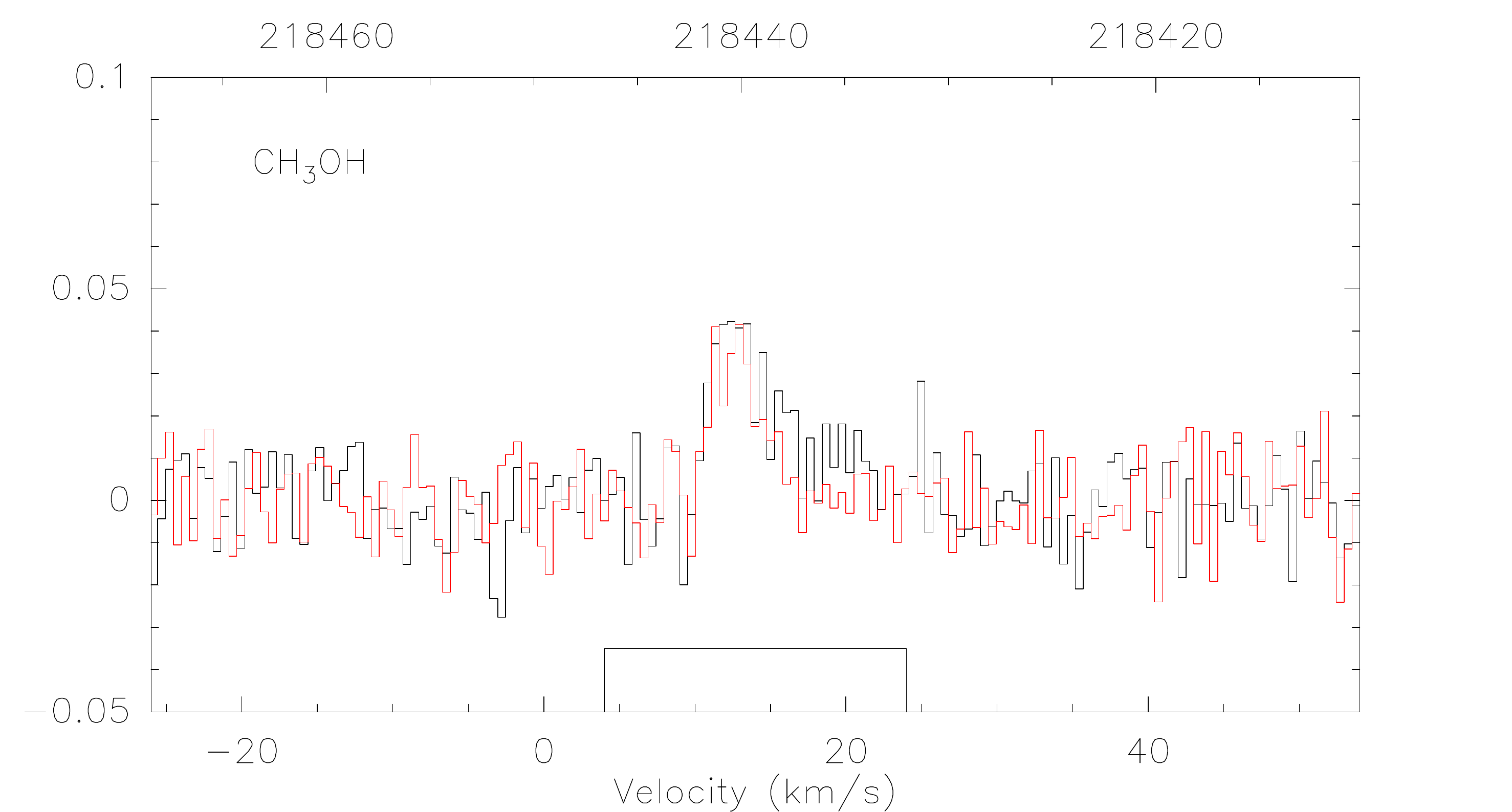} \\
\includegraphics[clip,width=0.5\columnwidth]{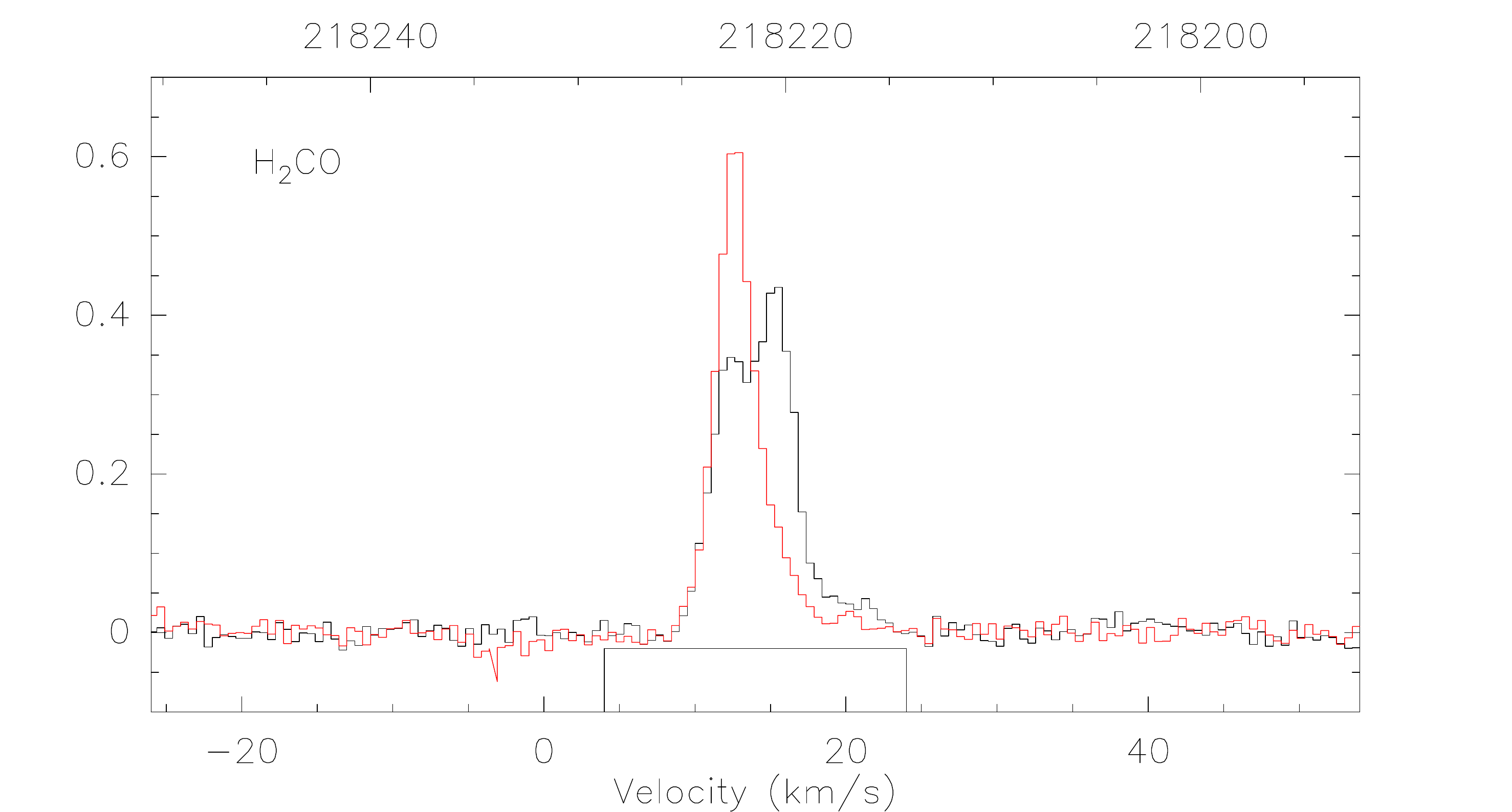} &
\includegraphics[clip,width=0.5\columnwidth]{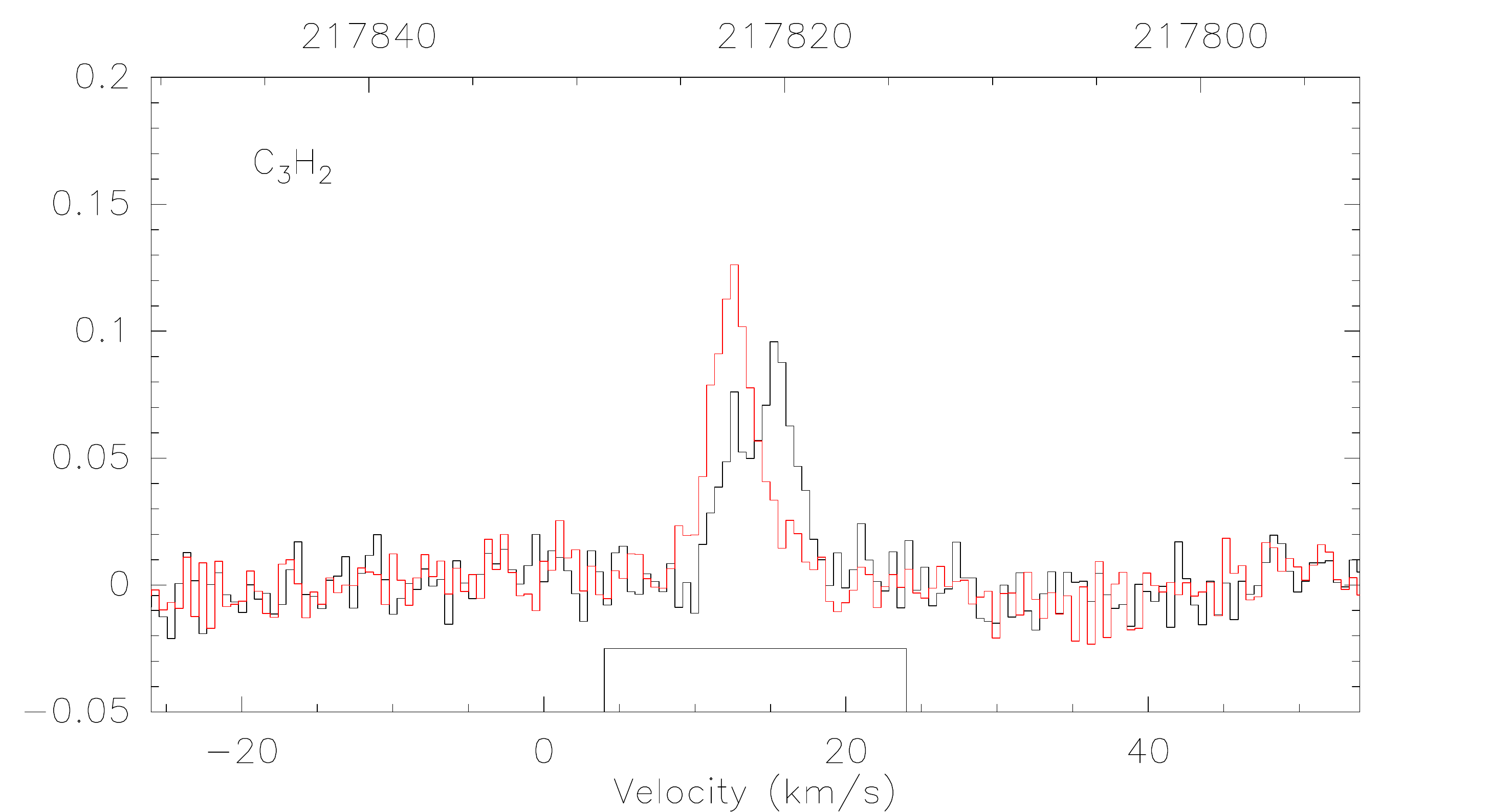} \\
\includegraphics[clip,width=0.5\columnwidth]{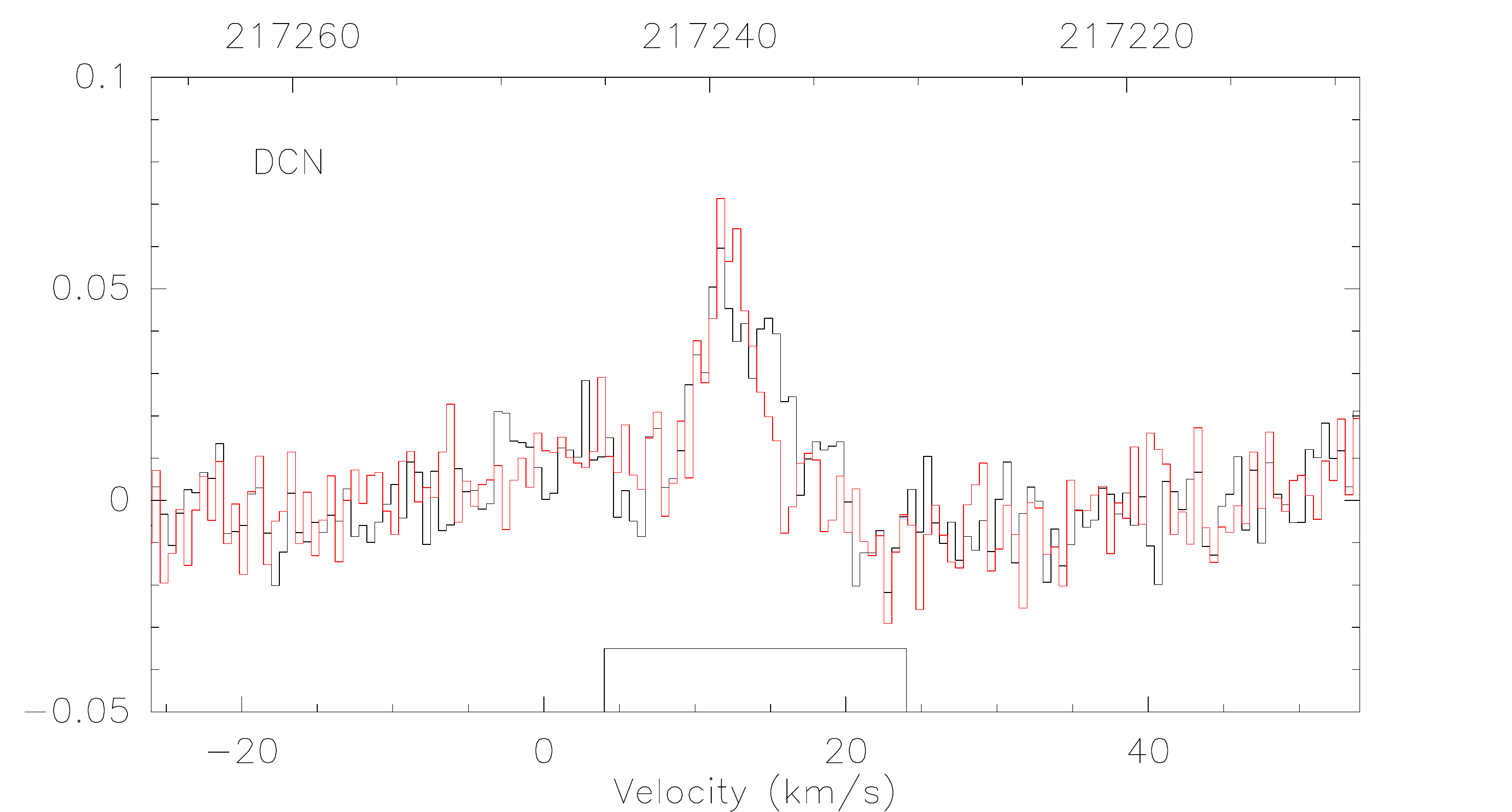} &
\includegraphics[clip,width=0.45\columnwidth]{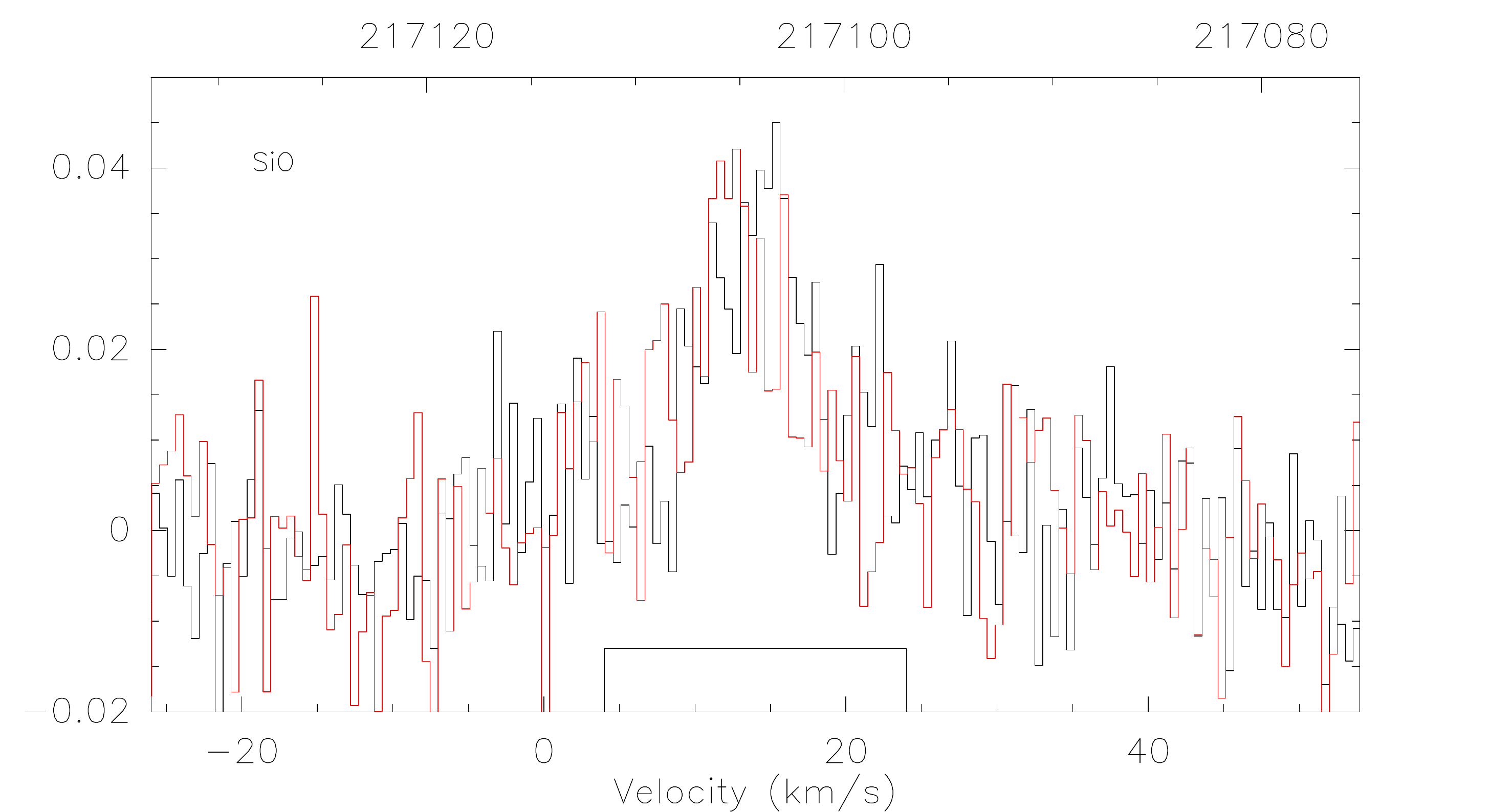}
\end{tabular}
\end{center}
\caption{APEX SHFI spectra (setting 1) of source \cn (black)
and source \cc (red) in antenna temperature (Kelvin) versus km/s. 
The beam FWHM is 32". All spectra smoothed with 5-channel box.  
The rectangles indicate data excluded from the baseline fit.
}
\label{spectra1}
\end{figure}
\begin{figure}
\begin{center}
\begin{tabular}{ccc}
\includegraphics[clip,width=0.5\columnwidth]{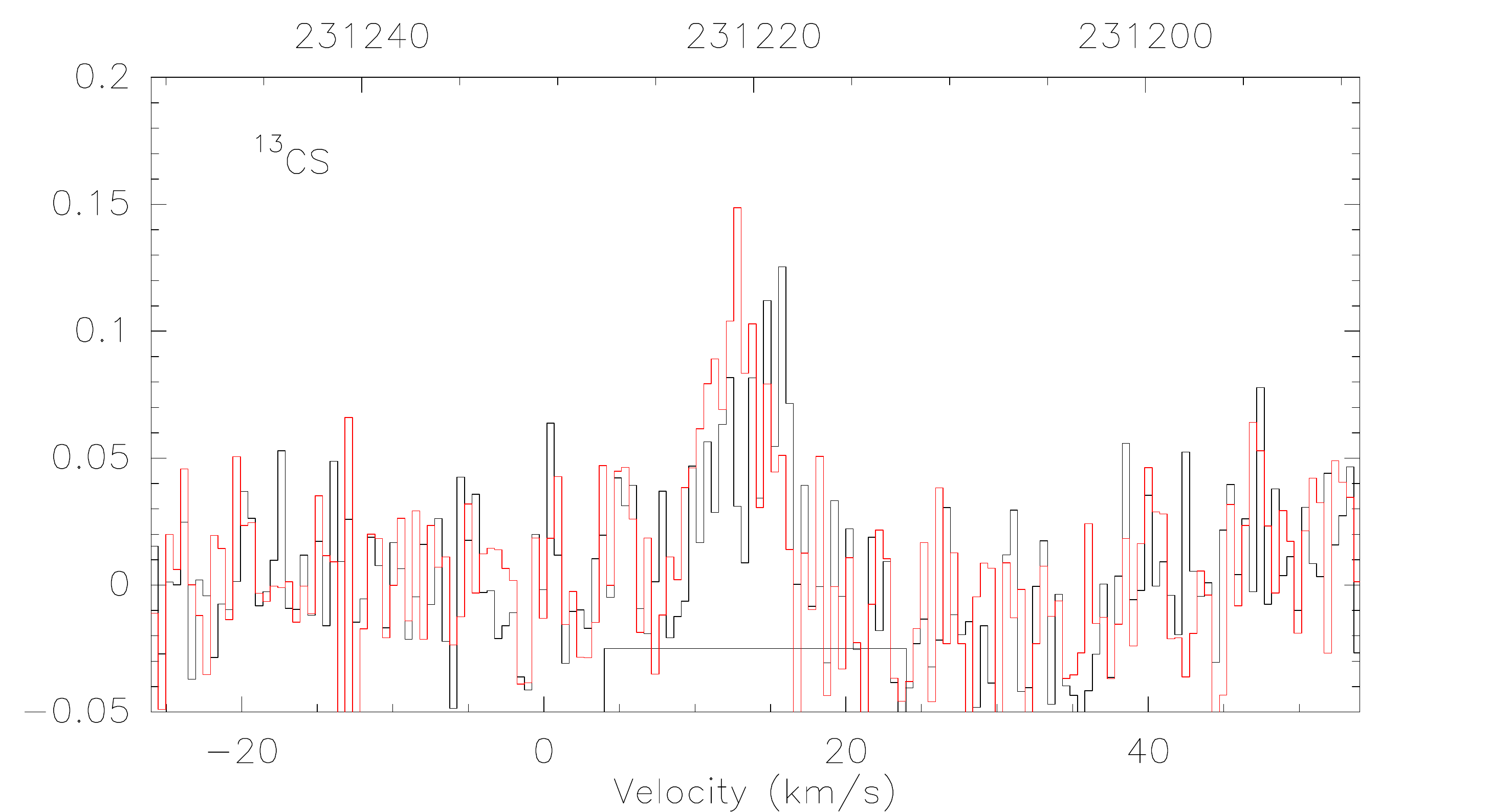} &
\includegraphics[clip,width=0.5\columnwidth]{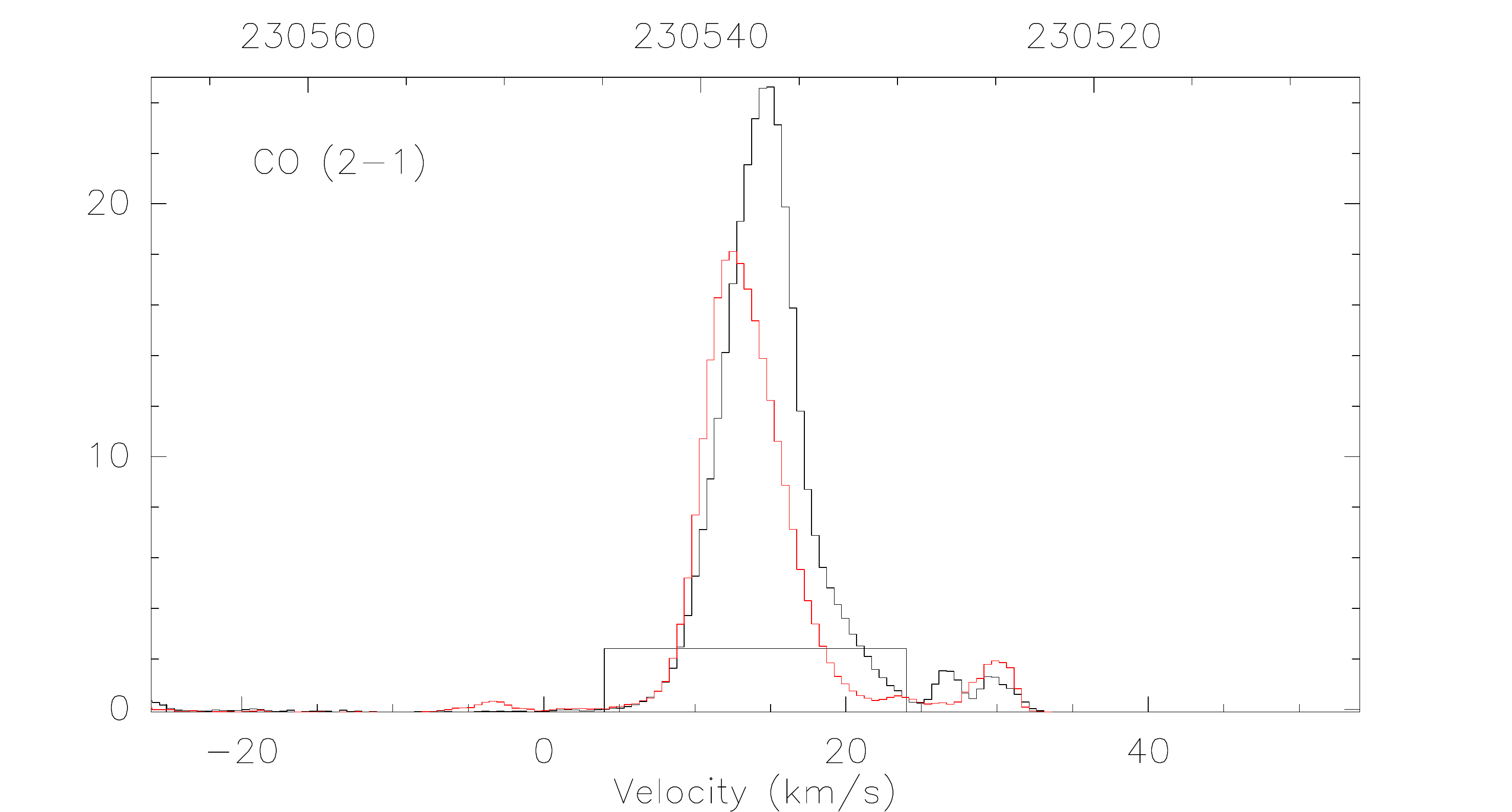} \\
\includegraphics[clip,width=0.5\columnwidth]{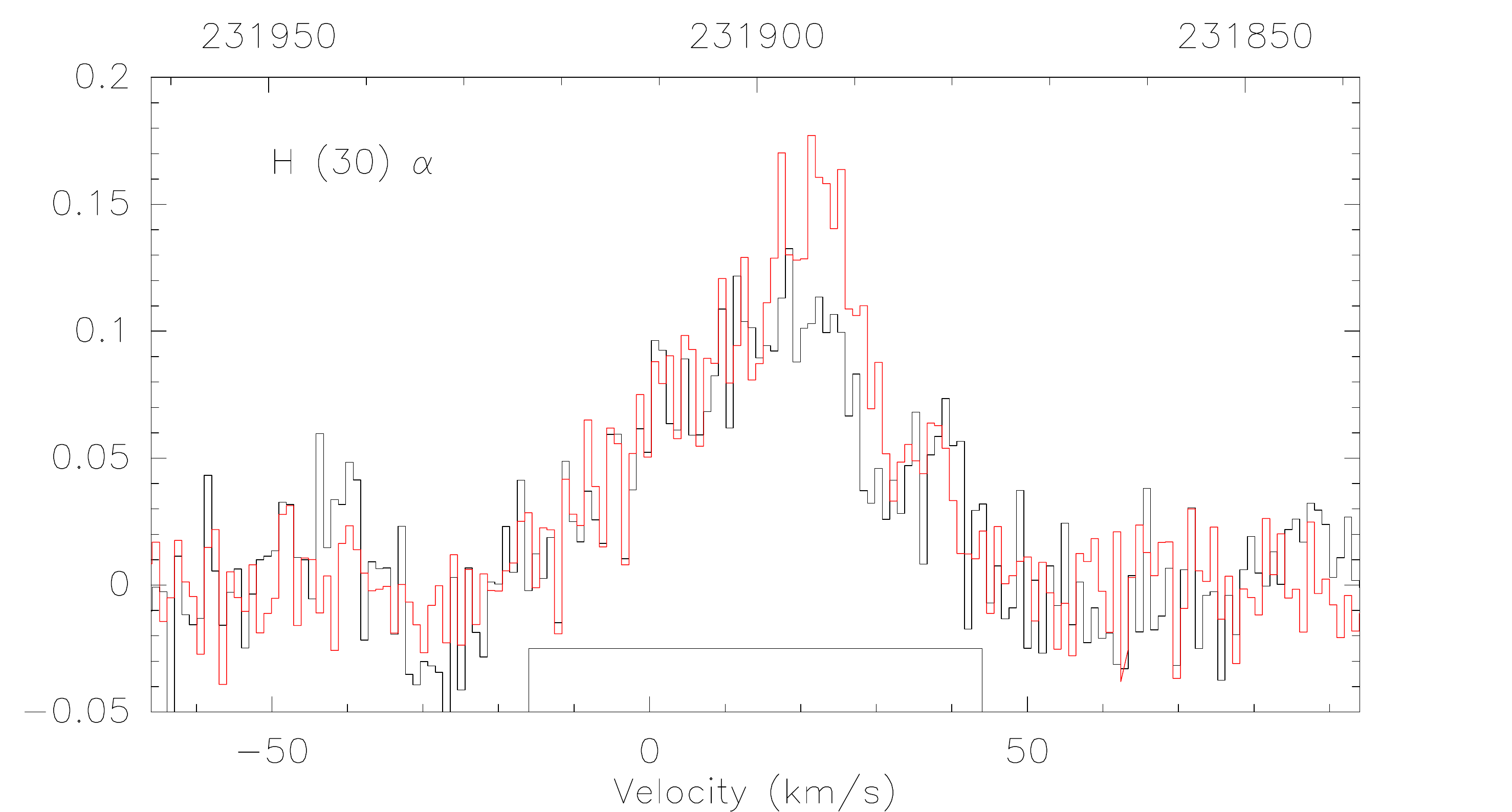} &
\end{tabular}
\end{center}
\caption{APEX SHFI spectra (setting 2) of source \cn (black)
and source \cc (red) in Kelvin versus km/s. The beam FWHM is 29". All
spectra smoothed with 5-channel box, 10-channel box for H$30\alpha$.
}
\label{spectra2}
\end{figure}
\begin{figure}
\begin{center}
\begin{tabular}{ccc}
\includegraphics[clip,width=0.5\columnwidth]{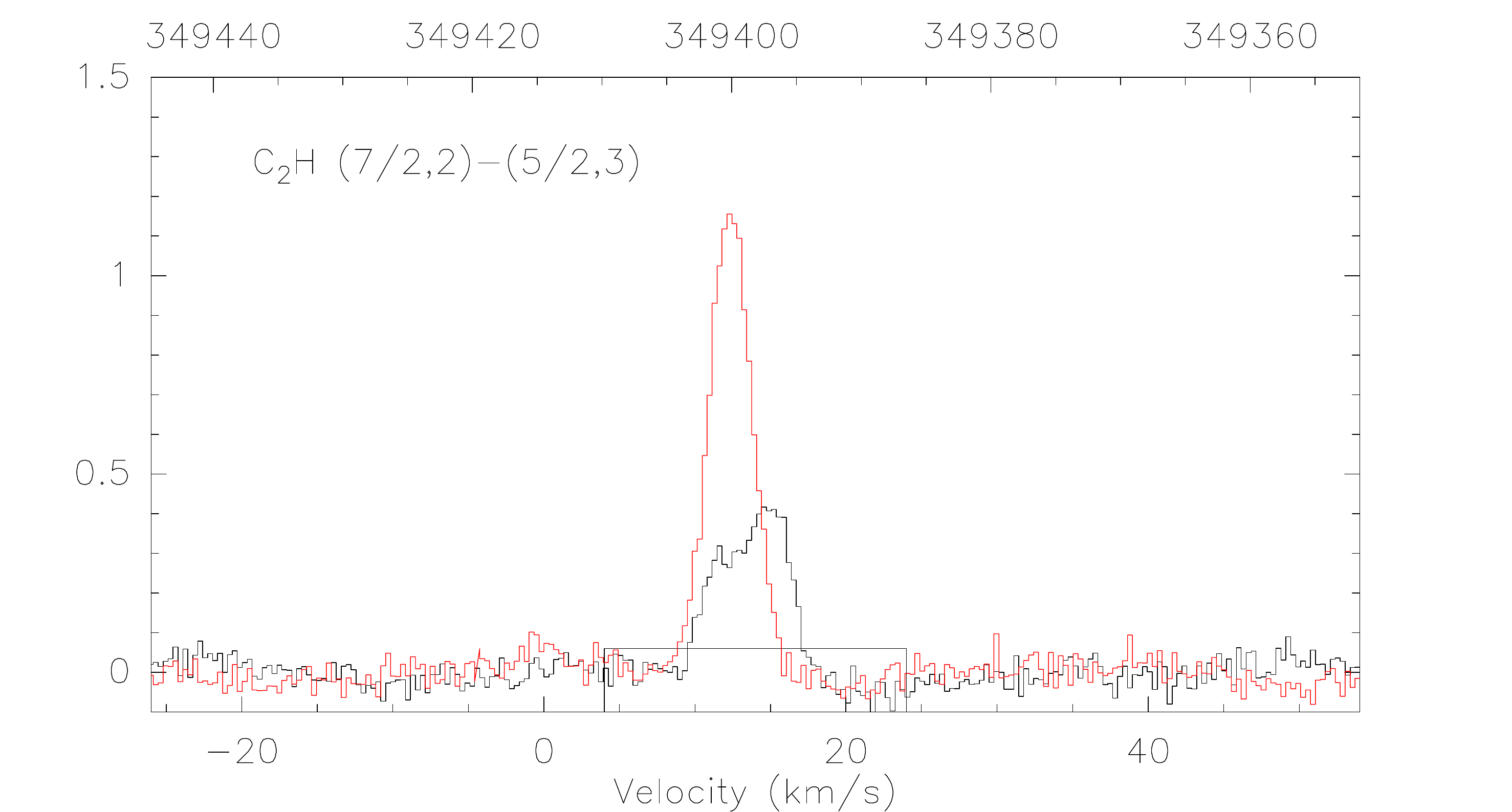} &
\includegraphics[clip,width=0.5\columnwidth]{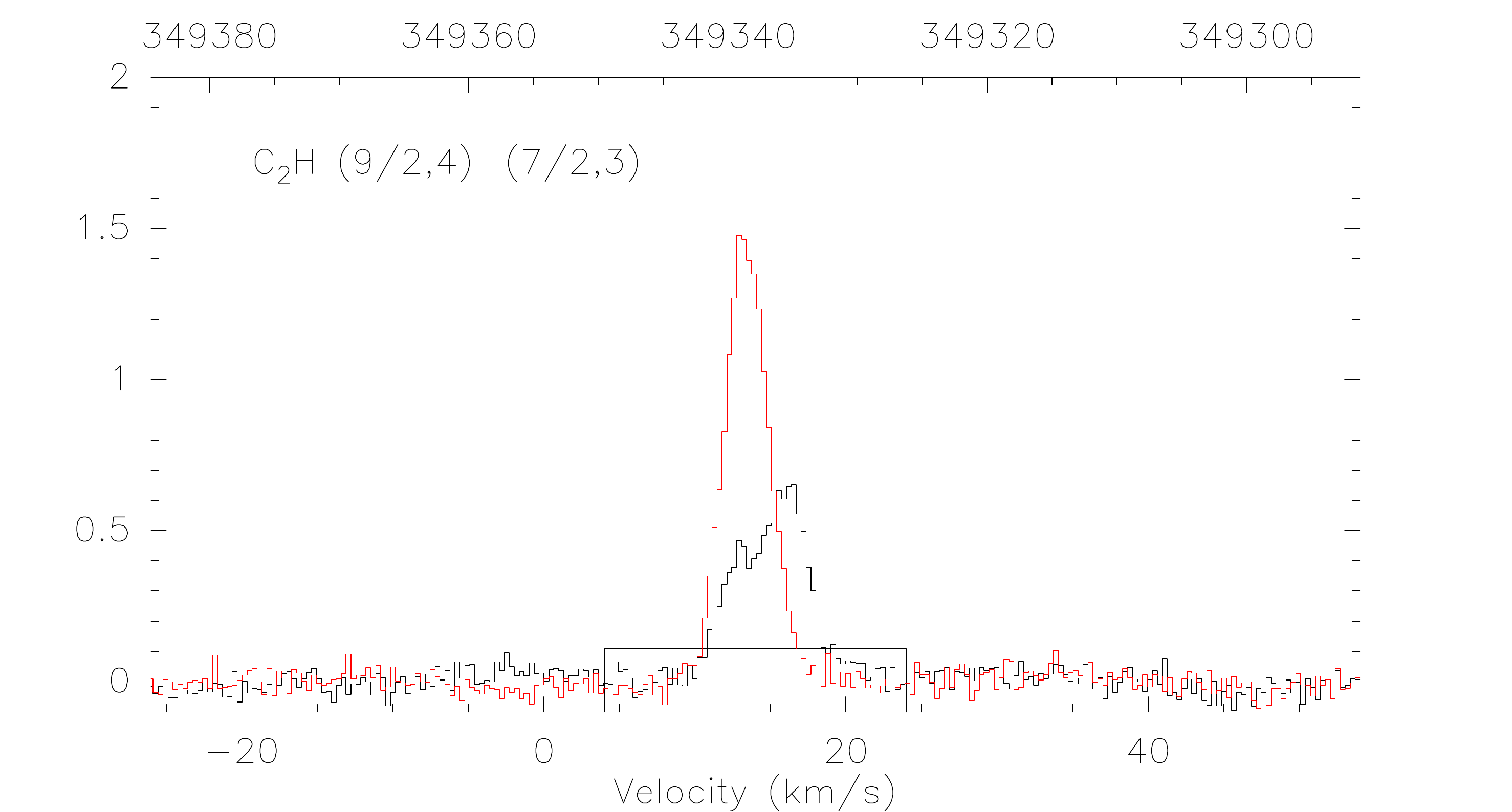} \\
\includegraphics[clip,width=0.5\columnwidth]{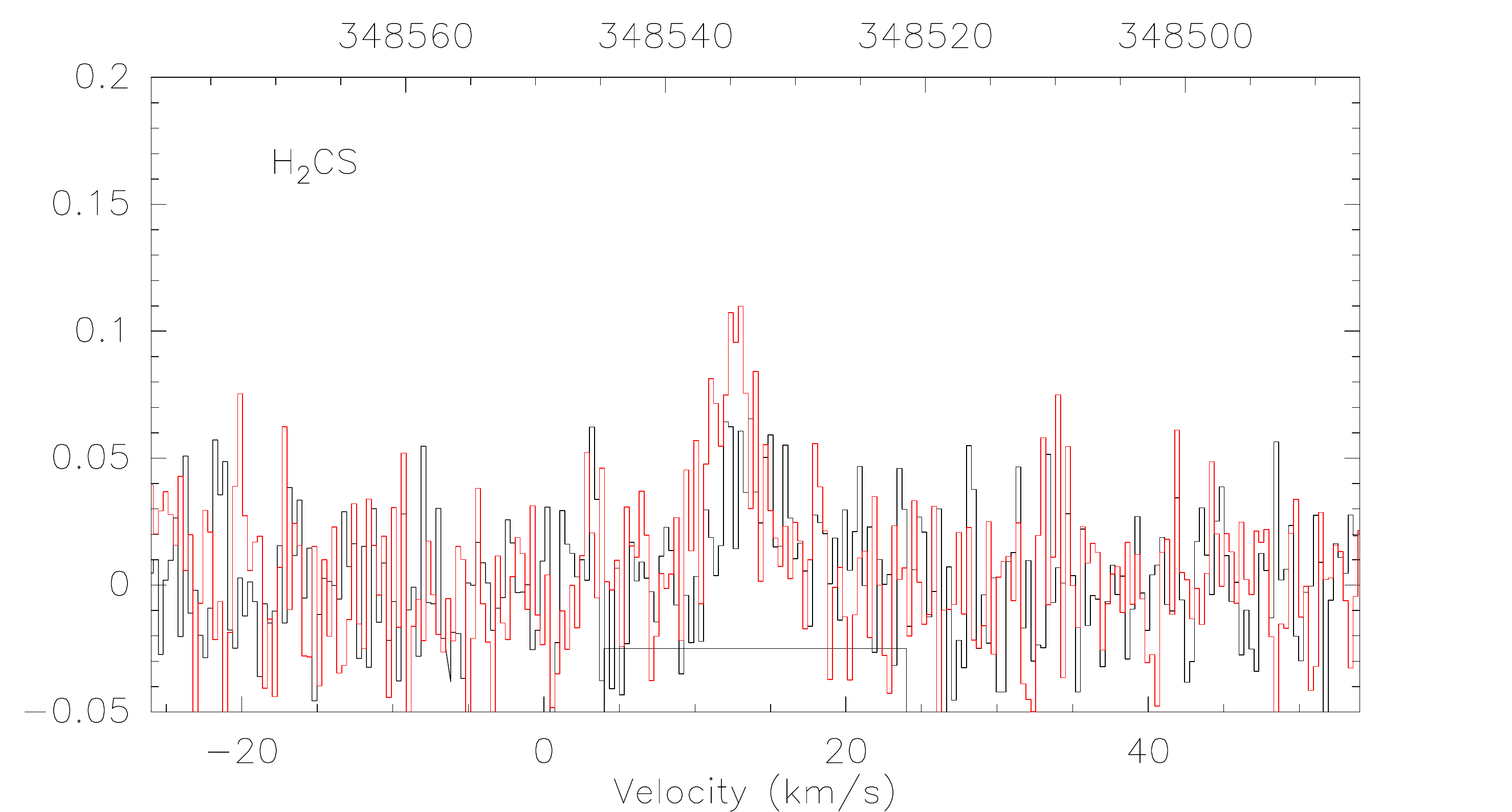} &
\includegraphics[clip,width=0.5\columnwidth]{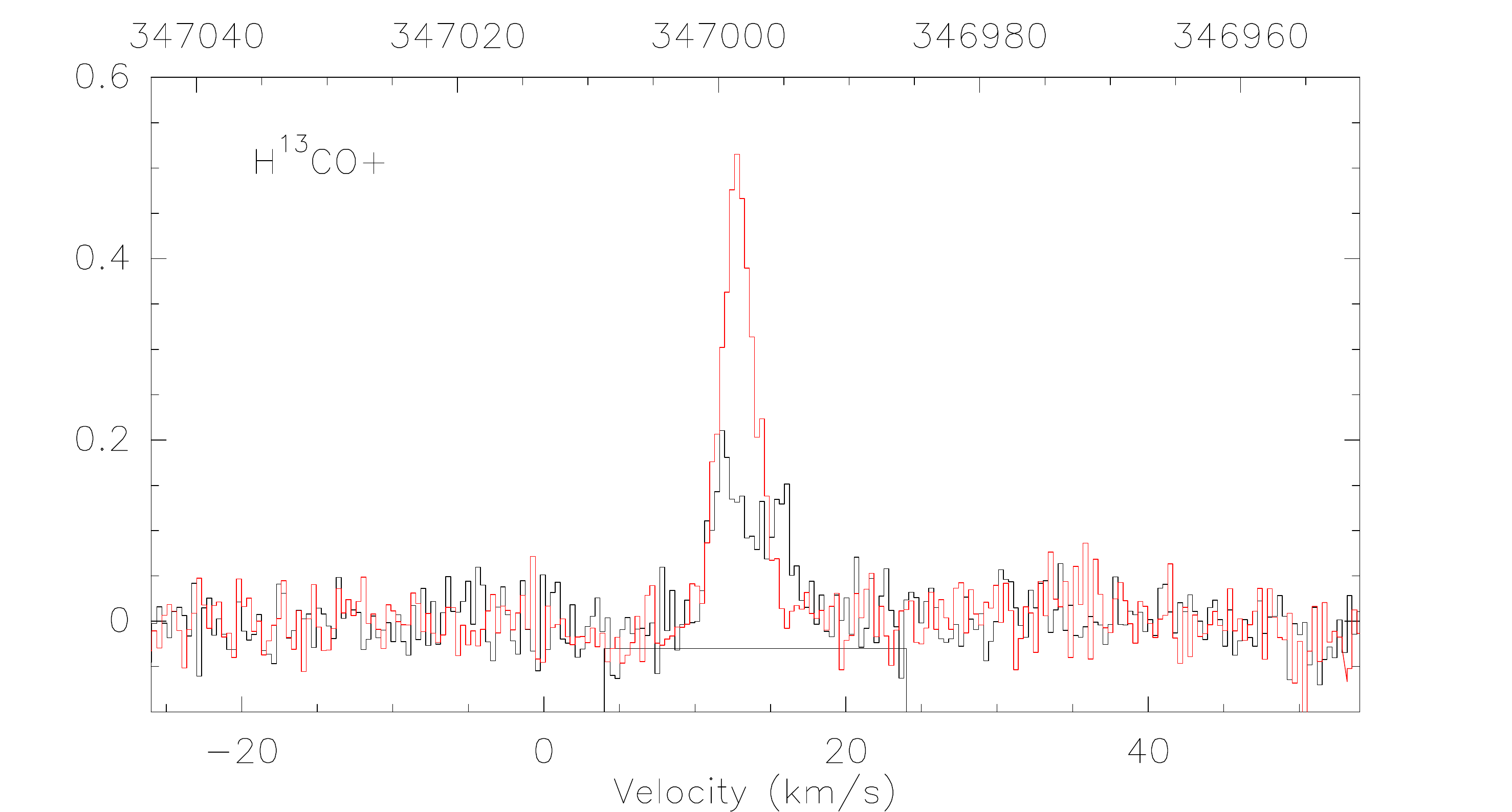} \\
\includegraphics[clip,width=0.5\columnwidth]{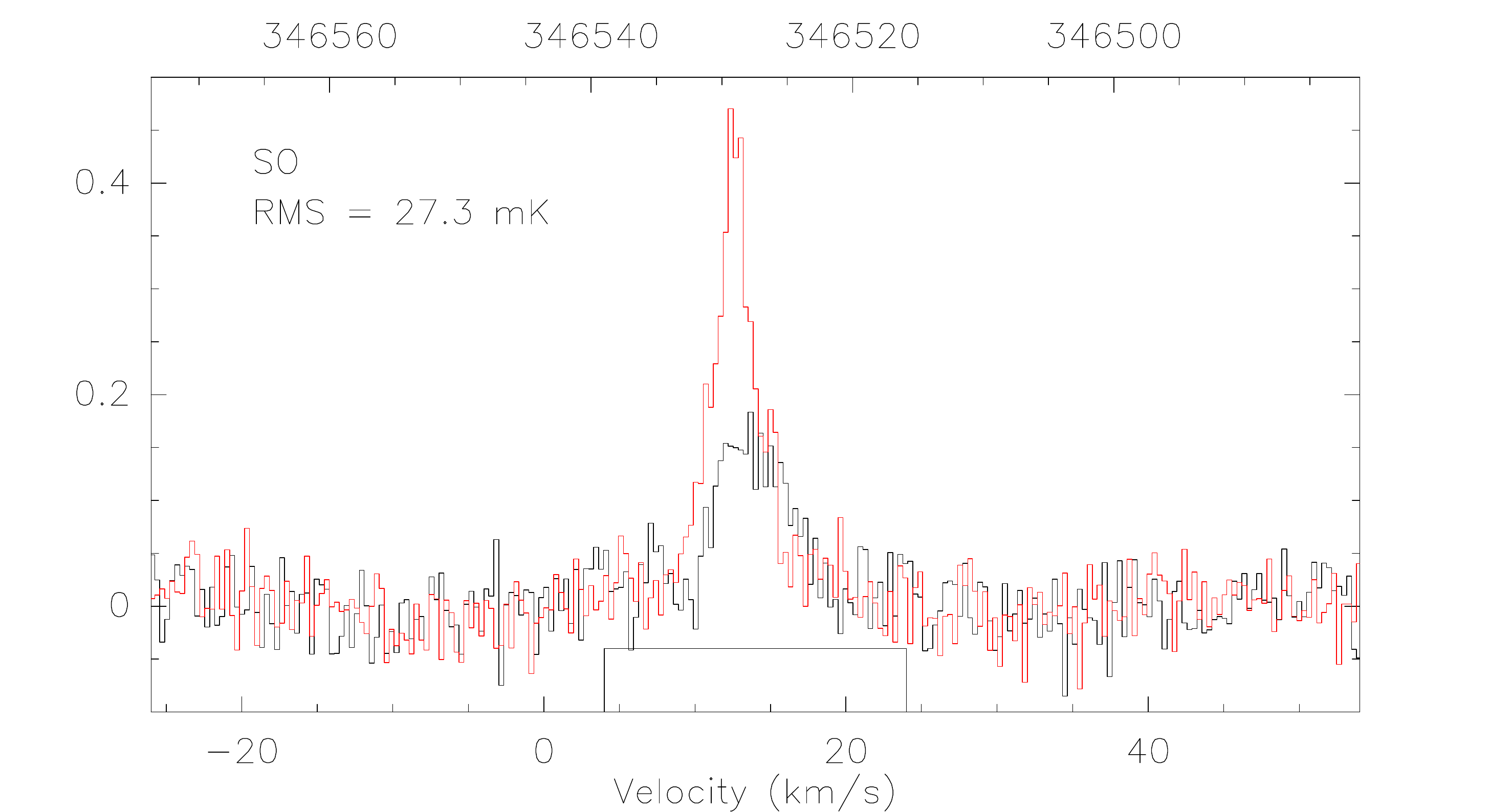} 
\end{tabular}
\end{center}
\caption{APEX SHFI spectra (setting 3) of source \cn (black)
and source \cc (red) in Kelvin versus km/s. The beam FWHM is 20". All
spectra smoothed with 5-channel box.
}
\label{spectra3}
\end{figure}
\begin{figure}
\begin{center}
\begin{tabular}{ccc}
\includegraphics[clip,width=0.5\columnwidth]{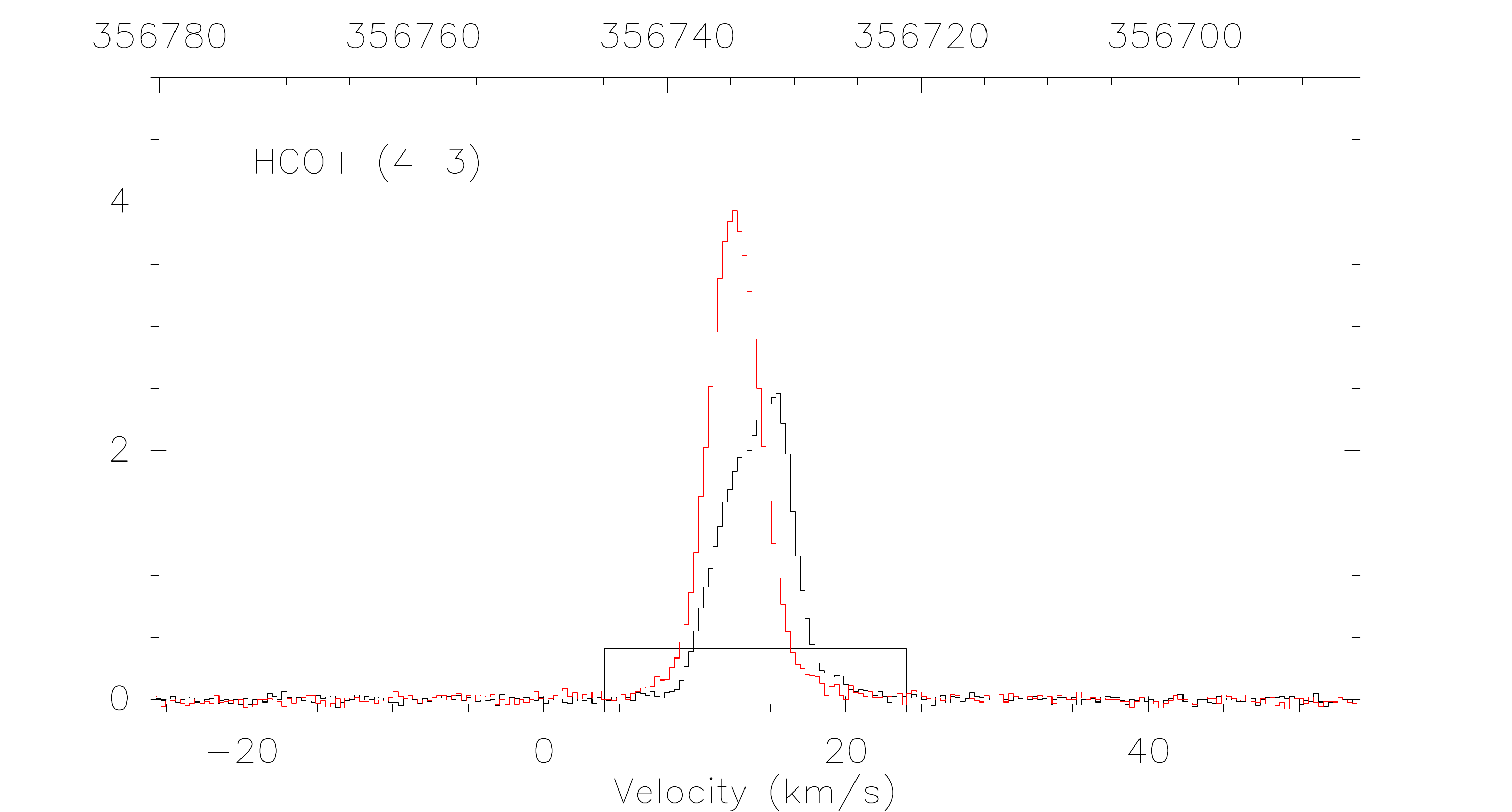} &
\includegraphics[clip,width=0.45\columnwidth]{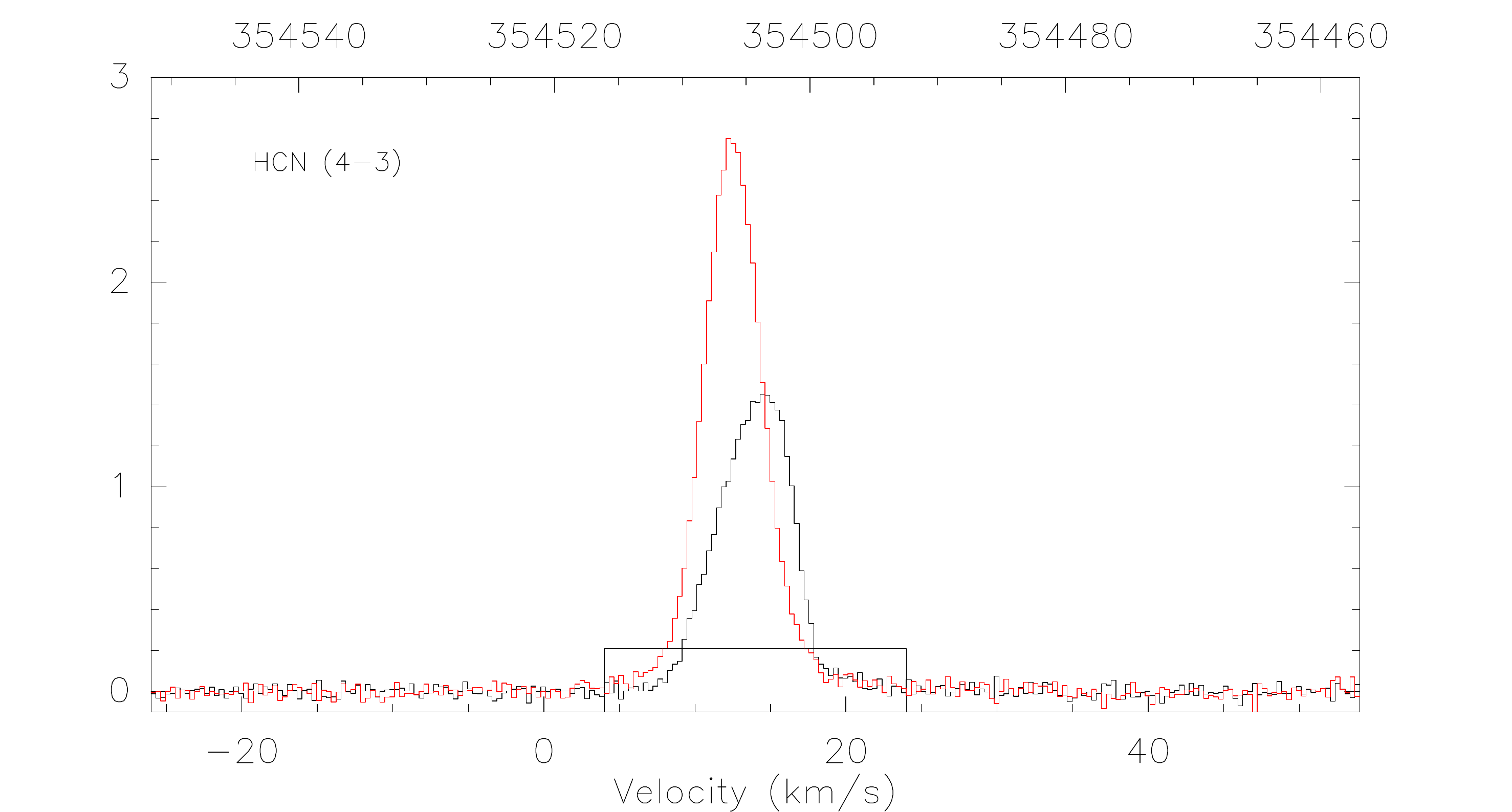}  \\
\includegraphics[clip,width=0.5\columnwidth]{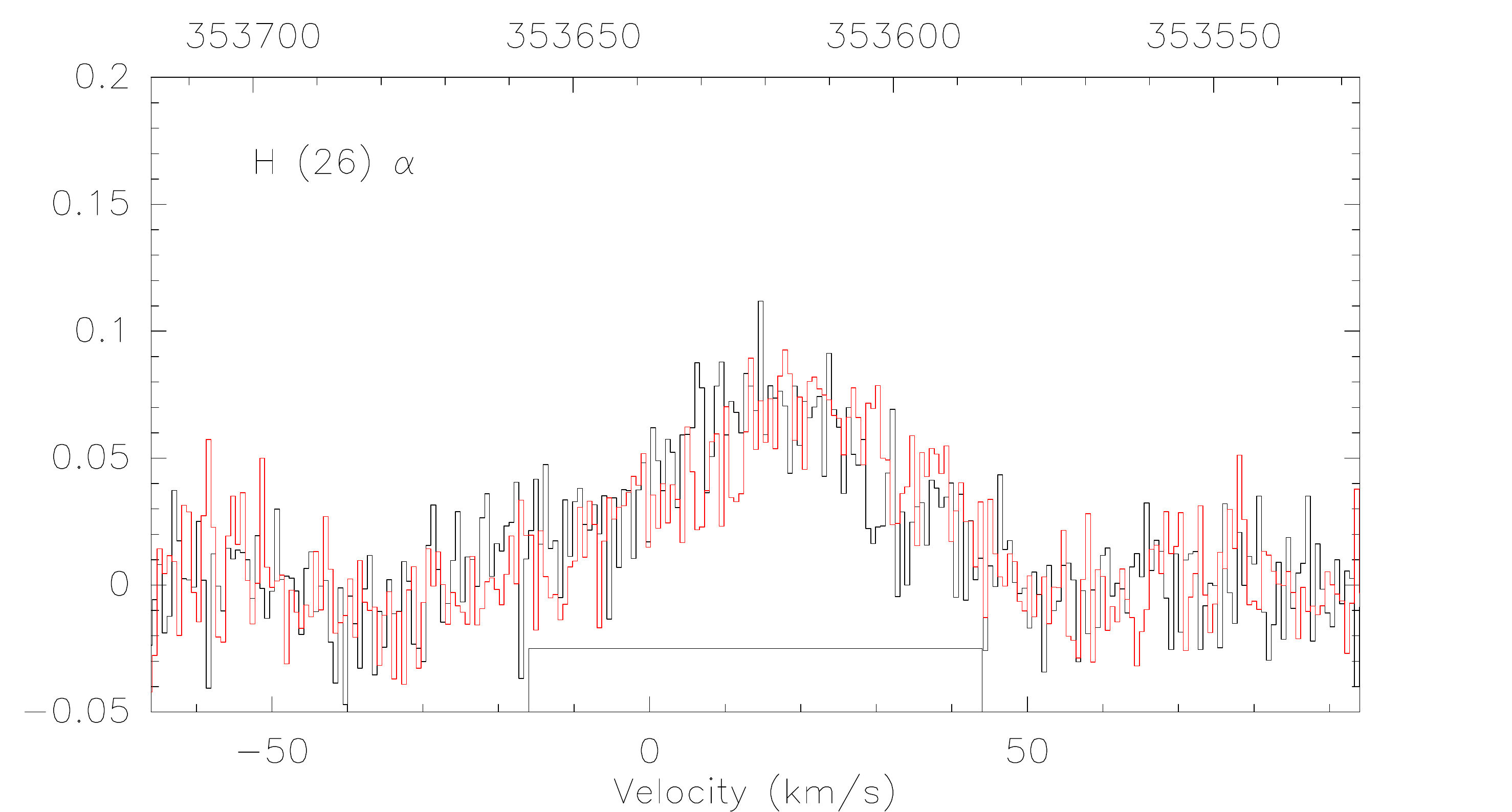}
\end{tabular}
\end{center}
\caption{APEX SHFI spectra (setting 4) of source \cn (black)
and source \cc (red) in Kelvin versus km/s. The beam FWHM is 19". All
spectra smoothed with 5-channel box, 10-channel box for H$26\alpha$.
    }
\label{spectra4}
\end{figure}
\begin{figure*}
\begin{center}
\begin{tabular}{c}
\includegraphics[clip,width=2.0\columnwidth]{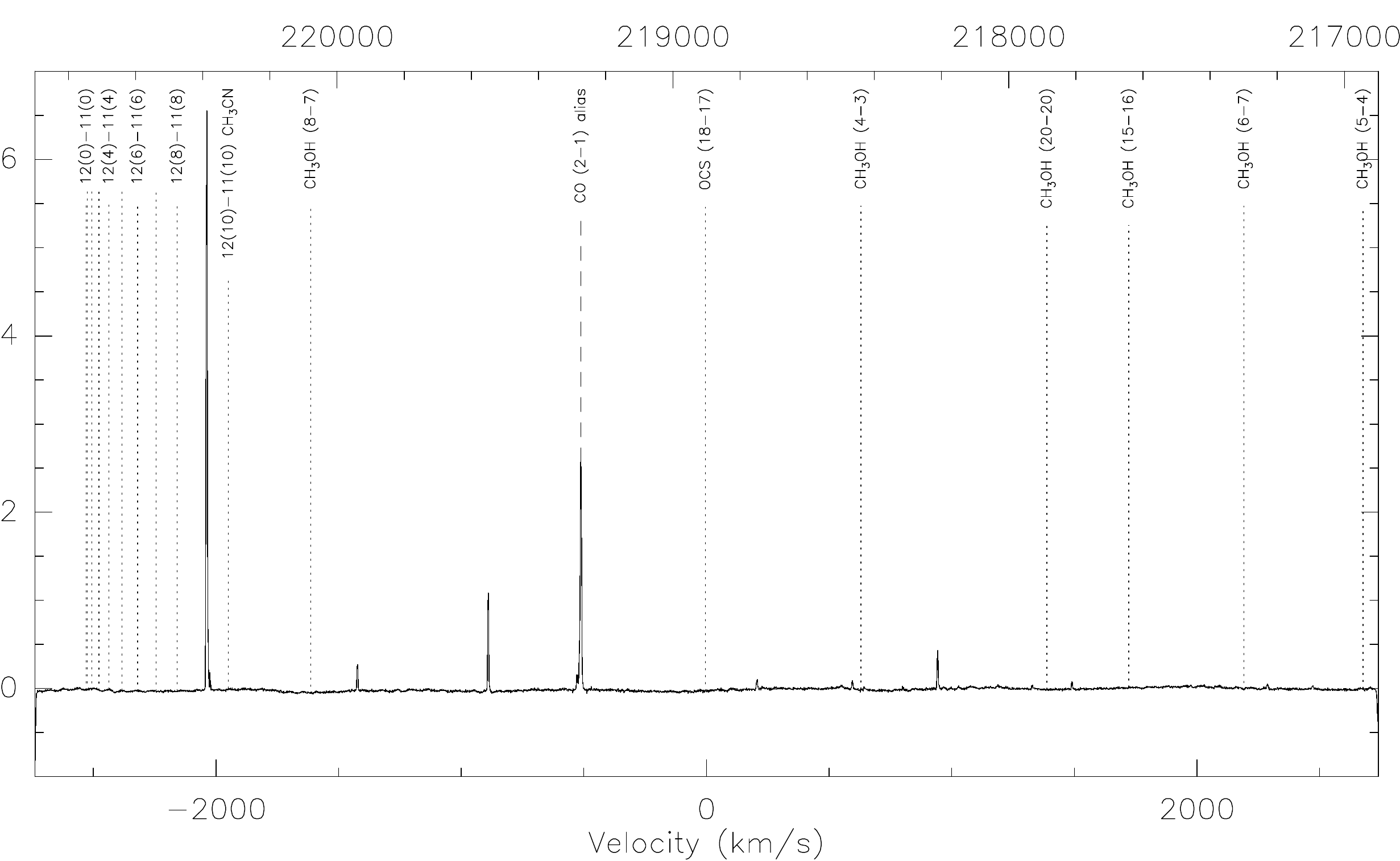} \\
\includegraphics[clip,width=2.0\columnwidth]{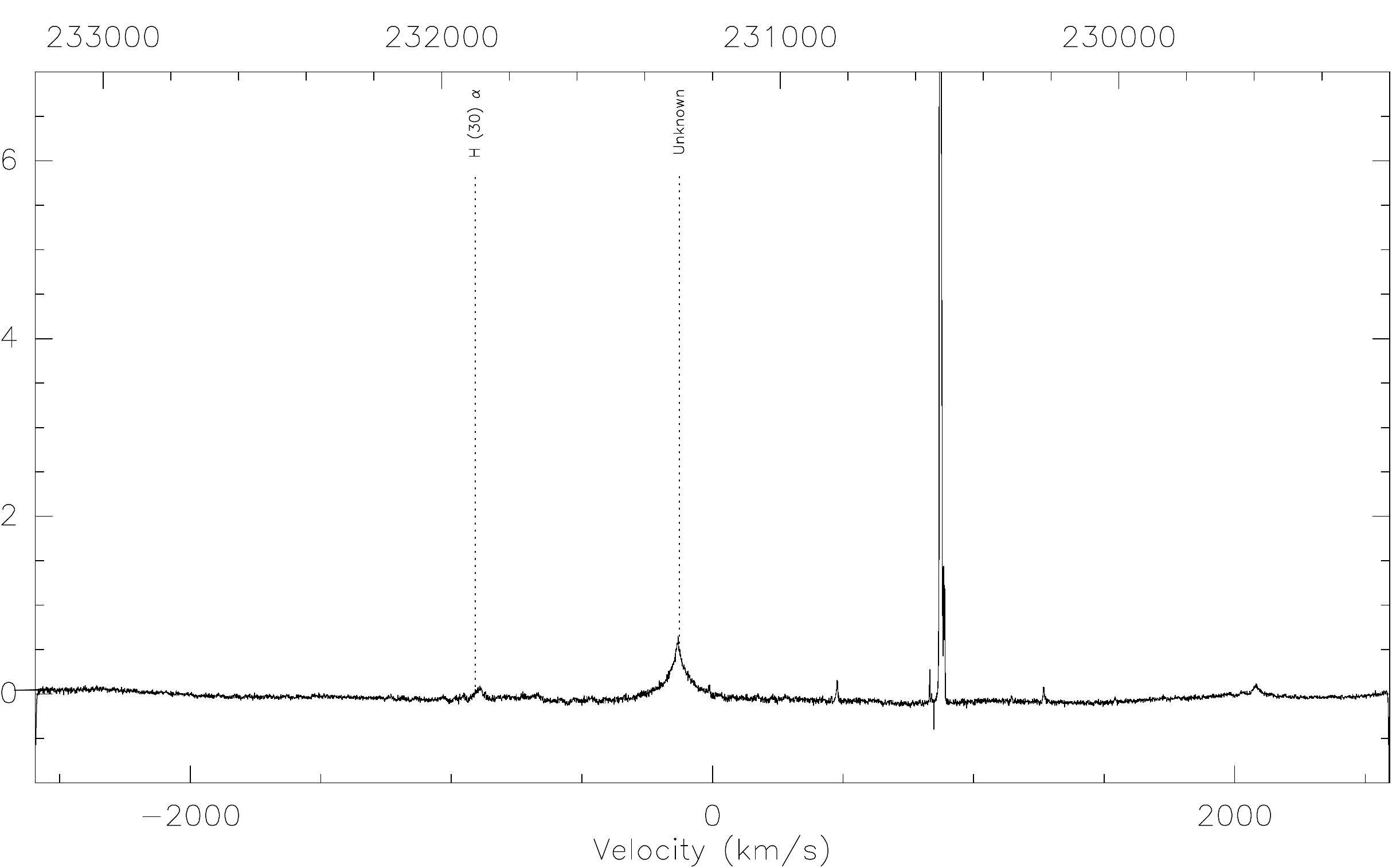} \\
\end{tabular}
\end{center}
\caption{Full-bandwidth spectra for settings 1 and 2 recorded with
SHFI on source \ca. All spectra were smoothed with a 10-channel
box. The ``wiggles'' of the baseline in the lower left panel are due
to instrumental problems.  Dotted lines mark frequencies of molecular
lines which we expected to see, but didn't.
    }
\label{spectrafull}
\end{figure*}
\begin{figure*}
\begin{center}
\begin{tabular}{c}
\includegraphics[clip,width=2.0\columnwidth]{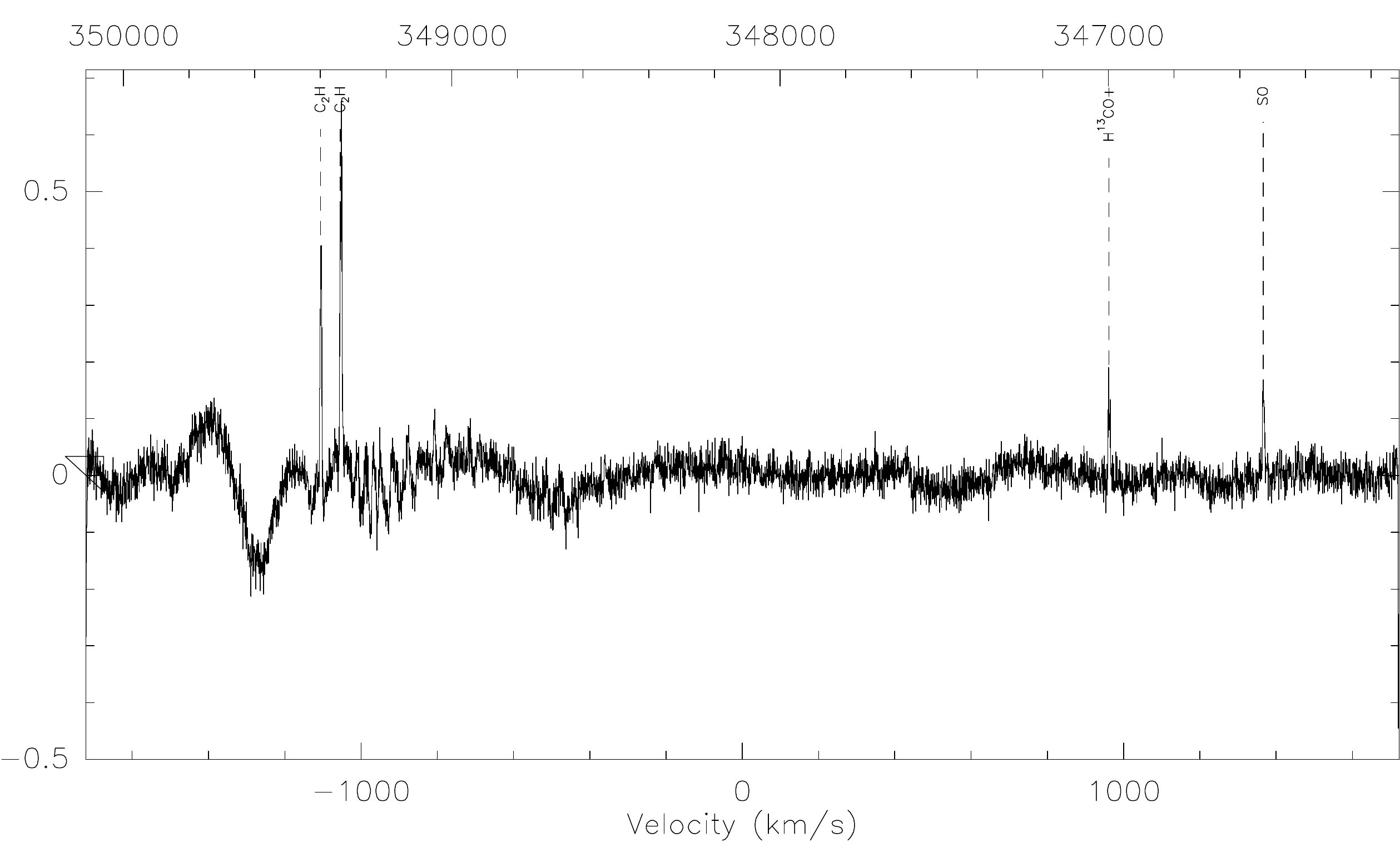} \\
\includegraphics[clip,width=2.0\columnwidth]{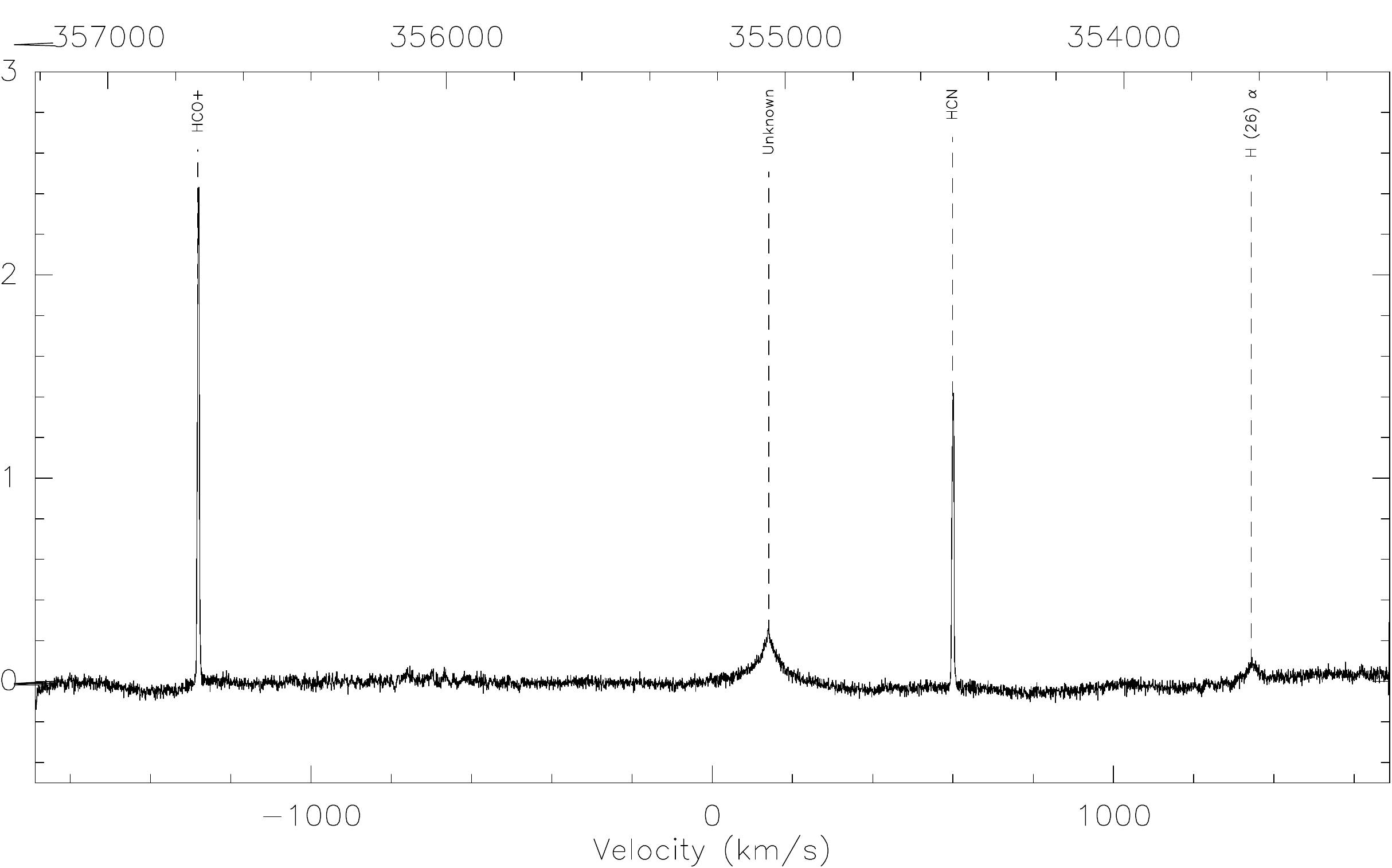} \\
\end{tabular}
\end{center}
\caption{Full-bandwidth spectra for settings 3 and 4 recorded with
SHFI on source \ca. All spectra were smoothed with a 10-channel
box. The ``wiggles'' of the baseline in the lower left panel are due
to instrumental problems.  Dotted lines mark frequencies of molecular
lines which we expected to see, but didn't.
    }
\label{spectrafull2}
\end{figure*}

The full spectra are shown in  Fig.~\ref{spectrafull} and
\ref{spectrafull2}, while spectra of the identified lines are shown in
Figs.~\ref{spectra1} to \ref{spectra4}.  In all plots, the units are
antenna temperatures in Kelvin versus km/s.

\begin{table}
\caption{Molecular lines (species, transition, frequency, and upper
energy level $E_u$) identified. All lines were detected in both positions.
Also the ratio of the antenna peak temperature between the offset position
(source \cc) and source \cn is given in the last
column (a hyphen indicates no significant difference for very weak
lines).}
\label{brightlines}
\centering
\begin{tabular}{cccccc}
\hline\hline
Species		& Transition		& $\nu_0$& $E_u$&$T_{\rm Sc}/T_{\rm S9}$\\
		&			& (GHz)		& (K)	&	\\
\hline
SiO		& 5-4			& 217.104984	& 31.2	& --	\\
DCN		& 3-2			& 217.238531	& 20.9	& 1.32	\\
c-HCCCH		& 6(1,6)-5(0,5)		& 217.822141	& 38.6	& 1.47	\\
CH$_3$OH	& 4(2,2)-3(1,2)		& 218.440047	& 45.5	& --	\\
H$_2$CO		& 3(0,3)-2(0,2)		& 218.222188	& 21.0	& 1.39	\\
\dots		& 3(2,2)-2(2,1)		& 218.475641	& 68.1	& 1.30	\\
\dots		& 3(2,1)-2(2,0)		& 218.760078	& 68.1	& 1.41	\\
C$^{18}$O	& 2-1			& 219.5603	& 15.8	& 1.19	\\
SO		& 5,6-4,5		& 219.949438	& 35.0	& 1.45	\\
$^{13}$CO	& 2-1			& 220.3986	& 15.86	& 0.97	\\
CO		& 2-1			& 230.538	& 16.6	& 0.75	\\
$^{13}$CS	& 5-4			& 231.220688	& 33.3  & --	\\
H$\alpha$	& H$30\alpha$	& 231.900930	& Rec.  & --	\\
SO		& 8,9-7,8		& 346.528594	& 78.8	& 2.40	\\
H$^{13}$CO$+$	& 4-3			& 346.998347	& 41.6	& 3.37	\\
H$_2$CS		& 10(1,9)-9(1,8)	& 348.534250	& 105.2	& --	\\
C$_2$H		& 4(7/2,4)-3(5/2,3)	& 349.399342	& 41.9	& 2.88	\\
\dots		& 4(9/2,4)-3(7/2,3)	& 349.339067	& 41.9	& 2.5	\\
H$\alpha$	& H$26\alpha$	& 353.622747	& Rec.  & --	\\
HCN		& 4-3			& 354.505469	& 42.5	& 1.80	\\
HCO$+$		& 4-3			& 356.734250	& 42.8	& 1.62	\\
\hline
\end{tabular}
\end{table}

\section{Results} \label{results}

\subsection{Cores and filaments}

We resolved the molecular cloud clump MM2 into numerous sources
(``compact cores'') both in the CS(2--1) line and sub-mm continuum
emission.  There is a clear correspondence of the compact cores seen at
mm-wavelength to their counterparts seen in the sub-mm.  The emission
has the general appearance of clumpy filaments.

The CS(2--1) line emission in NGC~3603 MM2 ranges from at least
10 to 20\kms.  The mean CS(2-1) velocity field shows basically two
filaments a few \kms apart, which was seen in CS spectra already by
\citet{2002A&A...394..253N}. One extends from source \cg  over \cn, \ca,
and  \cf at a systemic velocity of 15 --16 \kms (see Fig.~\ref{atca_img_cs}),
while the other one extends from source \cn over \ca to \ce at a systemic
velocity of about 12 \kms. As the two filaments overlap at the position
of source \cn, the line profiles are double-peaked here, also indicating
optically thin emission.

We observed the optically thin/thick pair of lines H$^{13}$CO$+$/HCO$+$
(Figs.~\ref{spectra3}/\ref{spectra4}) which can be used to trace infall 
\citep{1996ApJ...465L.133M,2007ApJ...663.1092K,2010ApJ...710..150C} 
if the optically thick line of HCO$+$ is double peaked, with a stronger
blue peak. The optically thin line of H$^{13}$CO$+$ is used to rule
out possible self-absorption (the line would have a single peak at the
frequency of the absorption).  Instead, we see that the H$^{13}$CO$+$
line is double peaked itself, and thus the line shapes do not indicate
infall, but are again the result of superposed filaments at different
LSR velocities.

Weak C$^{34}$S(2--1) emission from several of the compact cores was
detected as well (see Fig.~\ref{atca_img_c34s}) and appears to wrap
around the peak of the CS(2--1) emission of the strongest source \ca
(see Fig.~\ref{saboca} for the source nomenclature), a behaviour also
found by \citet{2009AJ....137..406B} from observations towards massive
warm molecular cores. The same is true for very weak emission peak at
the position of component \cn, while the C$^{34}$S(2--1) emission is
peaked at the positions of sources \cc and \ce. Similar offsets were
observed by \citet{2014A&A...572A..63I} in W33 Main (their Fig. A.6),
but no explanations for the offsets were given.

\subsection{Temperature of the compact cores}

Following \citet{1993ApJS...89..123M}, the gas temperature can
be determined from the peak flux ratio of selected formaldehyde
(H$_2$CO) lines. In particular, for two of the lines we detected
($3_{03}-2_{02}$/$3_{22}-2_{21}$), Fig.~13b of \citet{1993ApJS...89..123M}
indicates a range of temperatures (depending on the number density of
molecular Hydrogen ranging from $10^4$ to $10^5$ per cm$^3$) between
50 K and 60 K given a ratio of 5 for the peak flux ratio. The lower
value of this range is consistent with the temperature we fit (45 K)
to the formaldehyde line ratios using WEEDS (and for source \ca as
well). A similar gas temperature (47 K) has been determined for MM2 by
\citet{2011A&A...525A...8R}, while a dust temperature of 47 K for the MM2
pillar has been derived from SED fitting by \citet{2015ApJ...799..100D}.

\subsection{Mass of the compact cores}
\label{massofthecompactcores}

We computed the total gas and dust mass, $M_F$, of the cores based on
their 350 $\mu$m flux using equation D6 of \citet{2013ApJ...779..121G}
with an absorption coefficient of $\kappa_{\rm 850 GHz}=5.9$ cm$^2$/g
\citep[Table 1, column 5 of][] {1994A&A...291..943O}, giving, for
example, a mass of 250 to 330 M$_\odot$ for source \cn for the range of
temperatures given above. The results are listed in Table~\ref{sources}
(column 7) for all compact cores, assuming they have all the same
temperature of 50 K. The total mass of all compact cores in MM2 (therefore
not including source \cm) is about 1800 $M_\odot$ and therefore appears
to be consistent with the mass estimate of 1500 solar masses for MM2
by \citet{2002A&A...394..253N}. Adopting a value of $\kappa_{\rm
100 GHz}=0.2$ cm$^2$/g from an extrapolation using the data of
\citet{1994A&A...291..943O} to the wavelength of the ATCA observations,
the continuum RMS of 35 mJy/beam corresponds to more than 200 M$_\odot$
and thus explains why the cores were not detected in continuum emission
with ATCA.


We also computed, using the prescription of \citet{1988ApJ...333..821M},
the virial masses, $M_V$, of the four sources in Fig.~\ref{atca_cs} for
which we can measure the CS line widths. In the cases of a double-peaked
line profile (sources \cn and \ca), which are due to a superposition
of two filaments, we decomposed the profile into two and used the width
of the line component at the lower velocity (as they correspond better
to the velocity of the other two sources).  In these cases, the virial
masses would be underestimated and are indeed only about 30\% of the
flux-based masses, while the virial masses of the sources \cc and \ce
are about 50\% of the flux-based estimates. These results are also
listed in Table~\ref{sources}.

\subsection{Extended CO gas emission and outflow}

\begin{figure}
\begin{center}
\includegraphics[width=1.0\columnwidth]{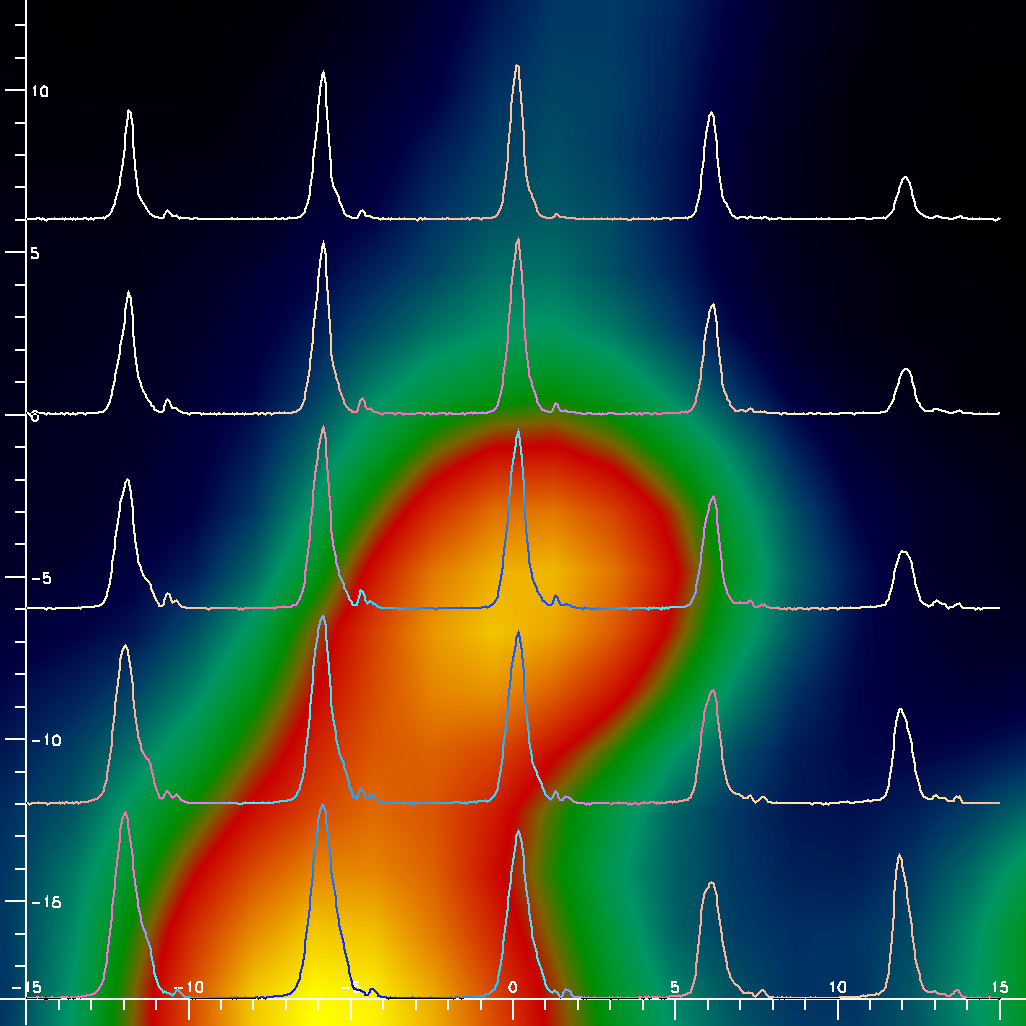}
\end{center}
\caption{APEX CO(3--2) (345 GHz) spectra in a grid of 25 positions (6"
spacing) in a region of MM2 centered on IRS 9A (source \cn).  North is
up and East to the left, and coordinates are in arc-seconds relative to
\ca. The individual spectra, each covering 60 km/s, are displayed
over the SABOCA map, with the center and base of every line aligned with
the corresponding pointing position. The maximum line peak is 28 K 
main beam temperature. The beam FWHM is 20".
}
\label{raster25}
\end{figure}

In Fig.~\ref{raster25} we show spectra of the CO(3--2) line at 25
positions forming a raster across MM2.  The decrease in line strength
at positions West of source \cn is consistent with there being no
CO emission (the beam FWHM is 20"). This is not the case in directions
North and East, indicating the presence of extended CO emission here.

Towards the South-East, a shoulder emerges in the line redshifted by a few
\kms. Comparing this location with the mean CS(2--1) velocity field
in Fig.~\ref{atca}, we conclude that the shoulder is due to the filament
associated with source \ca which has a line of sight (LOS) velocity of
about 16 \kms, while source \cn has a LOS velocity of about
12 \kms.

The CO (2-1) line shown in Fig.~\ref{spectra2} shows more emission
peaks at velocities of +25 -- 30 km/s (amplitudes of 2 K). These can
also be seen in Fig.~\ref{raster25}. Since they are separated from
the LSR velocity by more than the redshifted filament could explain,
we interpret this as a hint towards a high velocity outflow.  Further
evidence for an outflow comes from our detection of the (very weak) SiO
line (Fig.\ref{spectra1}), which is generally interpreted as an outflow
tracer \citep[e.g.][]{2007ApJ...663.1092K}.  Despite it's low SNR, the
shape of this line appears to be triangular, indicating the existence
of line wings.

\subsection{Radio-recombination lines}

\begin{figure}
\begin{center}
\includegraphics[width=1.0\columnwidth]{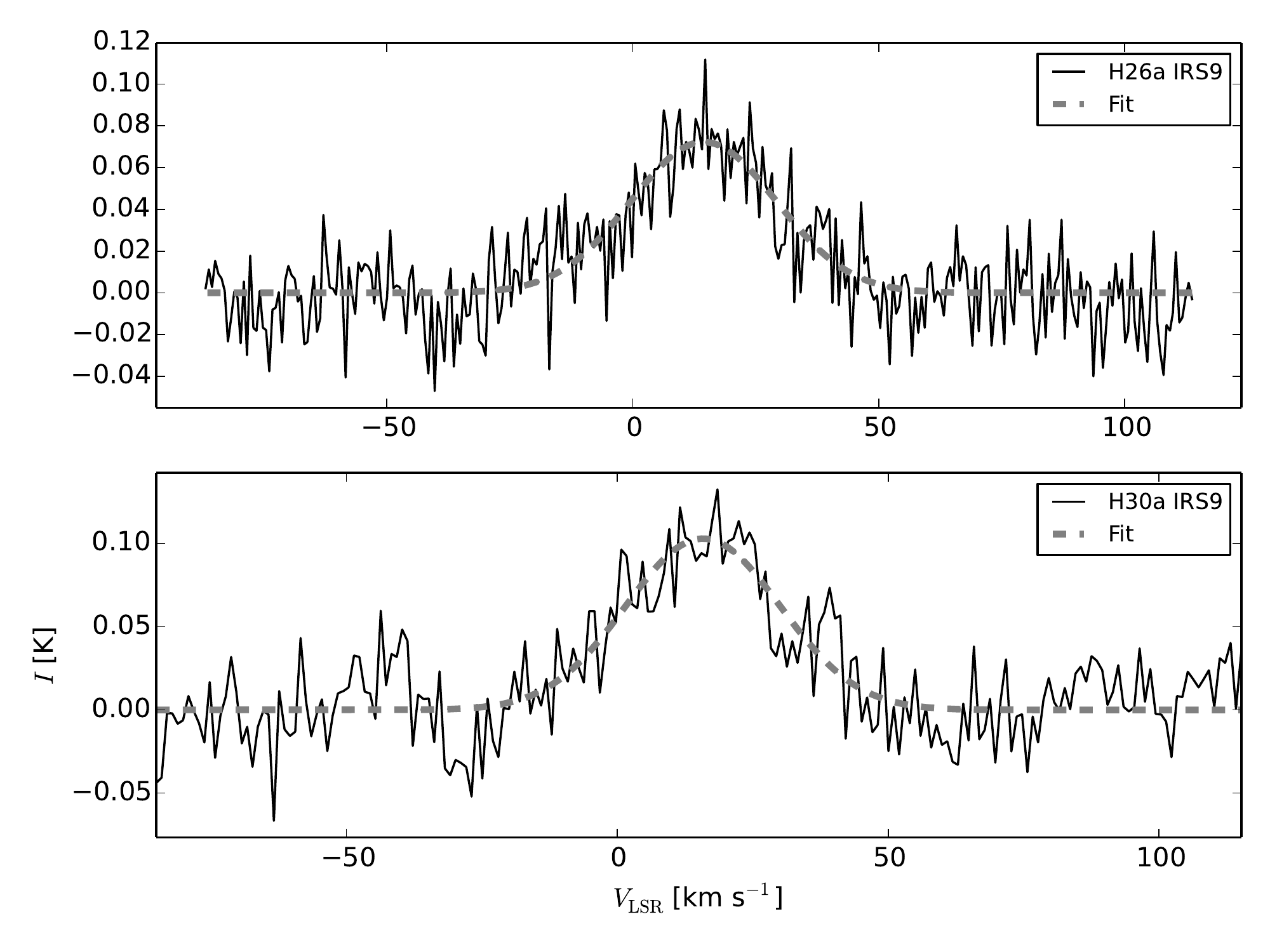}
\end{center}
\caption{APEX hydrogen recombination lines  H$26\alpha$ and H$30\alpha$
with Gaussian fits.
}
\label{rrl}
\end{figure}

Figure \ref{rrl} shows the millimeter hydrogen recombination line (RL)
emission towards the center of IRS 9. The detection of these lines
demonstrates the effect of photoionization feedback from the central
massive star(s) in IRS 9. The RL emission appears to be extended, since
the line is also detected away from IRS 9. High angular resolution
RL mapping is needed to determine the nature of the RL emission,
whether diffuse \citep[e.g.,][]{1998ApJ...501..710G}, or an ultracompact
or hypercompact \HII\ region \citep{2007prpl.conf..181H}.

The FWHMs of the H$26\alpha$ and H$30\alpha$ lines are the
same within uncertainties (FWHM$_\mathrm{H26} = 34.8 \pm 1.8$ km
s$^{-1}$, FWHM$_\mathrm{H30} = 33.5 \pm 2.0$ km s$^{-1}$).  This is
consistent with previous observations in other regions which show
that mm RLs are free from collisional broadening compared to cm RLs
\citep[e.g.][]{2008ApJ...672..423K,2012A&A...547L...3G}. Therefore,
the observed line width can be explained as a combination of the
thermal width of ionized gas at $10^4$ K ($\sim 20$ km s$^{-1}$)
and dynamical broadening due to bulk motions of order $2c_s$, where
$c_s \sim 10$ km s$^{-1}$ is the speed of sound of the ionized gas.
The centroid LSR velocities of both lines are also the same within
uncertainties, and consistent with the IRS 9 systemic velocity as
derived from dense molecular gas: $V_\mathrm{LSR,H26} = 14.4 \pm
0.8$ km s$^{-1}$, $V_\mathrm{LSR,H30} = 15.8 \pm 0.9$ km s$^{-1}$.
Finally, we note that the velocity-integrated line intensities
for both lines are about the same. For resolved observations, and
assuming LTE and low line and free-free continuum optical depths,
the line ratio should scale approximately linearly with frequency
\citep[e.g.][]{2008ApJ...672..423K,2012A&A...547L...3G}.  Two possible
explanations for our observations are that: the H$26\alpha$ line, because
its lower optical depth, has a smaller filling factor within the APEX
beam than the H$30\alpha$ line; or that the H$30\alpha$ line is amplified
due to non-LTE effects \citep[e.g.][]{2013ApJ...764L...4J}. Sub-arcsecond
angular resolution RL observations are necessary to test these hypotheses.

\section{Discussion} \label{discussion}

The most obvious feature of our sub-mm spectra of sources
\cn and \ca is the lack of lines typically seen in hot-cores
\citep[e.g.][]{1993A&A...276..489O}, especially the series of lines of
Methylcyanide, Methanol, and OCS in the 217 -- 221 GHz band (see upper
left panel of Fig.~\ref{spectrafull}).

A prominent feature instead is the presence two Ethynyl lines near
349 GHz.  According to \citet{2008ApJ...675L..33B}, these lines are
seen in all evolutionary stages of massive stars beginning with infrared
dark clouds (IRDCs), and continuing via high-mass protostellar objects
(HMPOs) to ultracompact \HII\ regions (UCHII). When comparing our
full spectrum for setting 3 (Fig.~\ref{spectrafull}) to Fig.~1 of
\citet{2008ApJ...675L..33B}, it is obvious that source \cn is not a
HMPO. The mean Ethynyl line widths are 3.0 \kms for the two earlier
evolutionary stages, while they are 5.5 \kms for the UCHIIs. On source
\cc, we measured a line width of 3.3 \kms, and on \cn 6.0 \kms. However,
the latter is clearly a superposition of two line profiles originating
in different filaments, and fitting double-Gaussian profiles yield an
average width of $3.1 \pm 0.4$ km/s. Therefore, we would classify these
compact cores as IRDCs even though their temperature is about twice
that one would expect for IRDCs \citep{2006A&A...450..569P}.  As shown
in Fig.~\ref{atca6}, none of the compact cores shows free-free emission
at 5 GHz, thus do not appear to harbor UCHII regions.

\begin{figure}
\begin{center}
\includegraphics[width=0.8\columnwidth]{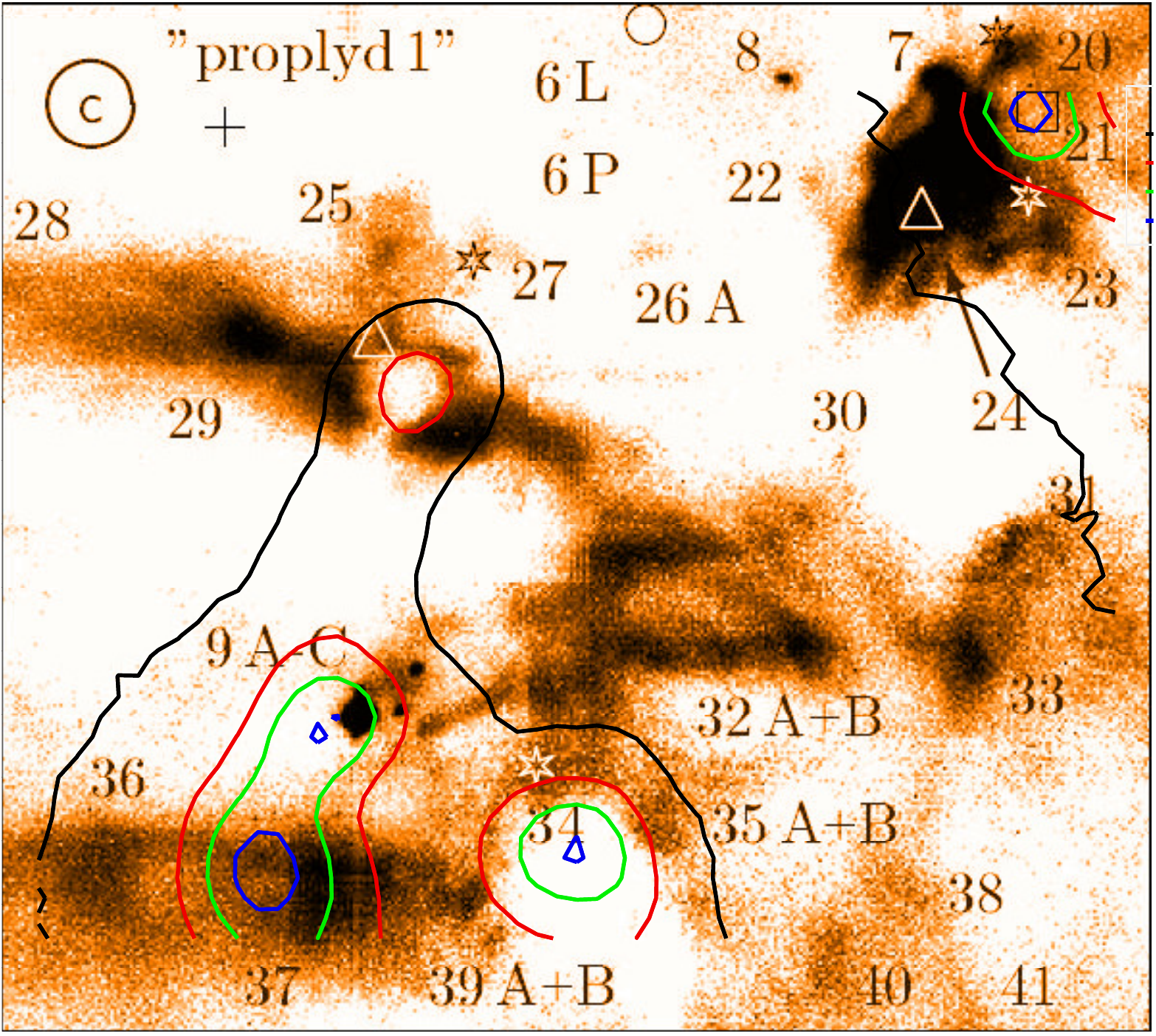}
\end{center}
\caption{TIMMI2 11.9 $\mu$m image \citep[Fig. 4c of
][]{2003A&A...400..223N} with overlaid contours of the SABOCA
map. The image size is 95" by 110". The labeling is from
\citet{2003A&A...400..223N} and denotes the position of the
cluster with an open circle, the positions of IR sources from
\citet{1977ApJ...213..723F} with asterisks, while all other labels refer
to sources given by \citet{2003A&A...400..223N}. The photo-center of
the SABOCA source closest to IRS 9A is shifted by about 3" to
the East-South-East.  Component \cm of the SABOCA image coincides
quite well with a maser source indicated by the square, according to
\citet{2003A&A...400..223N}.
}
\label{timmi2}
\end{figure}

\begin{figure}
\begin{center}
\includegraphics[width=1.0\columnwidth]{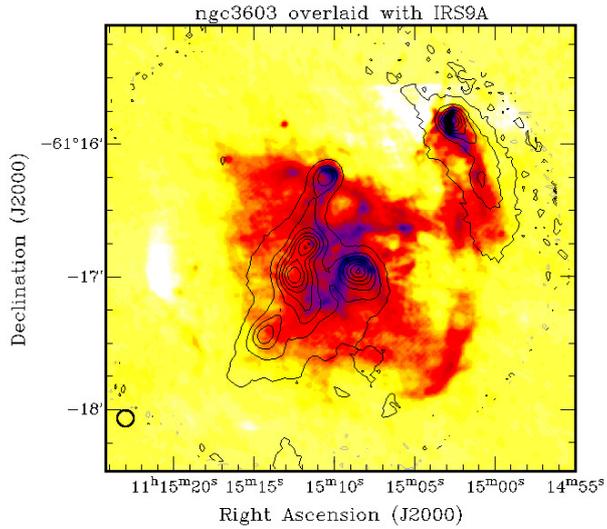}
\end{center}
\caption{Overlay of SABOCA contours on a 5 GHz image of the \HII\ region
obtained from ATCA data by \citet{2002ApJ...571..366M}. The blue colored
image regions correspond to the largest 5 GHz fluxes, while the yellow
color denotes the faintest 5 GHz emission. The contour levels are at 
87\%, 73\%, 60\%, 46\%, 33\%, 19\%, and 06\% of the maximum. The ATCA beam
is 2", the SABOCA beam is 7.8" and is shown in the lower left corner.
}
\label{atca6}
\end{figure}

If we look at the MIR emission in the region around IRS 9A
\citep{2003A&A...400..223N}) and overlay the contours of the 350 $\mu$m
map we obtained with SABOCA (Fig.~\ref{timmi2}), we see that sources
\cc, \cg (tip of the eastern pillar), and \cm seem opaque to background
MIR emission, consistent with a classification of IRDCs. IRS 9A itself
is quite close to source \cn, but apparently in front of it. (The
MIR position of \citet{2003A&A...404..255N} is about 3.6" away from the
position of \cn, even though more recent astrometric analysis by Brandner
(priv. comm.) reduces the offset to 1.8".) Similarly, extended MIR
emission seen towards \ca must originate in the forground while still
associated with warm dust in the region. It thus appears that IRS 9A
may have been formed more recently than the OB cluster stars out of a
compact core like the ones we found in MM2.

\begin{figure}
\centering
\begin{tabular}{l}
\includegraphics[width=1.0\columnwidth]{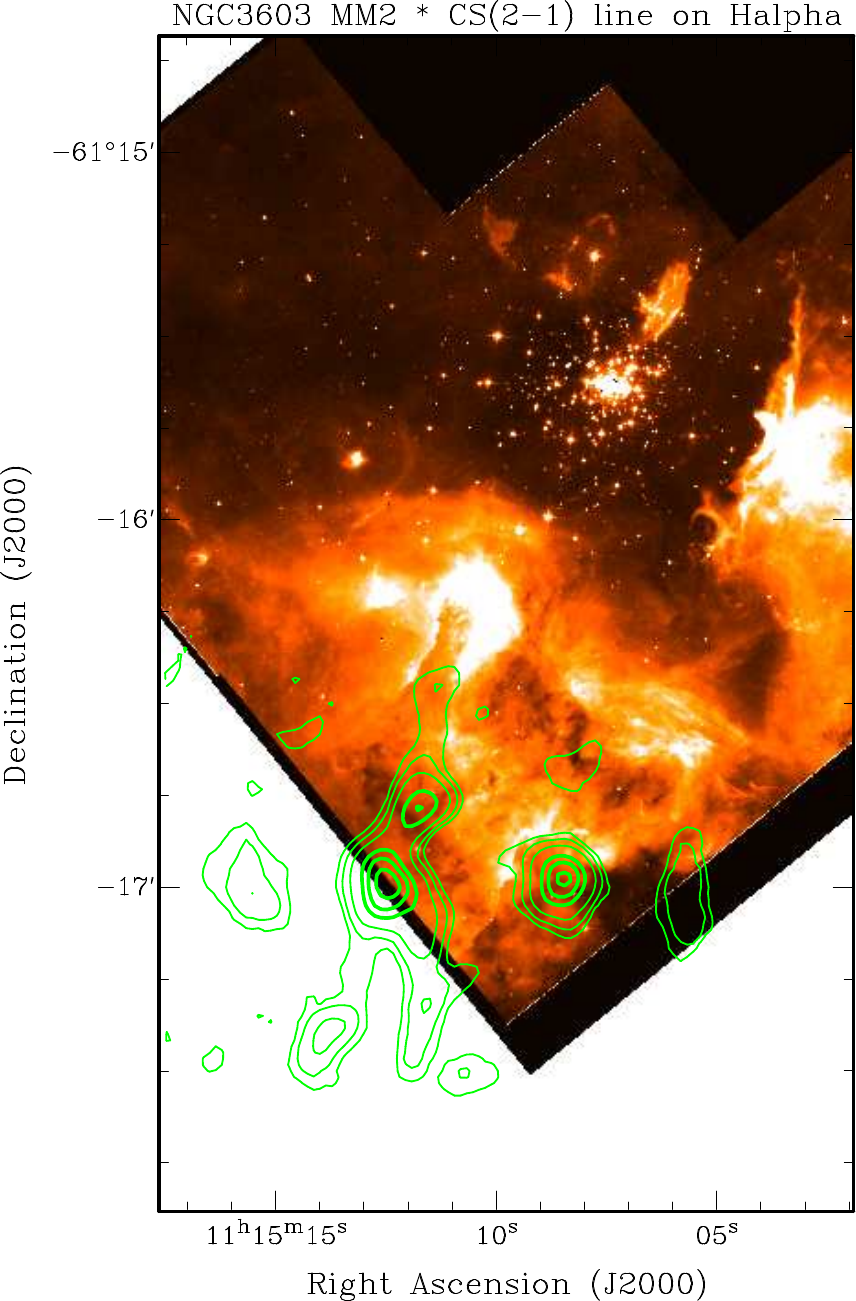}
\end{tabular}
\caption{ATCA CS (2--1) contour overlay on an HST H-$\alpha$ image 
\citep{2000AJ....119..292B}.
}
\label{mm2hst}
\end{figure}

Evidence for ionized gas around IRS 9A has been found from Spitzer
spectra, showing lines of [Ne\,{\footnotesize II}] and [S\,{\footnotesize
IV}] \citep{2008ApJ...680..398L}. Similarly, \citet{2011A&A...525A...8R}
reached the conclusion that ``the cluster is strongly interacting with the
ambient molecular cloud'' based on their map of the FUV radiation field.
Both Figs.~\ref{atca6} and \ref{mm2hst} illustrate this by showing strong
free-free emission at 6 cm and H$\alpha$ recombination line emission in
the NGC 3603 region at the head of the eastern (source \cg) and western
(source \cm) pillar, as well at the circumference of source \cc facing
the cluster. These regions likely contribute to the RL emission we
detected.

The fact that compact cores make up the clump MM2 and the close
association of the IR source 9A to one of them (source \cn)
may fit into the picture of sequential star formation from the
north to the south \citep{1999AJ....117.2902D}, away from the
OB cluster which itself is possibly the result of a cloud-cloud
collision \citep{2014ApJ...780...36F}.  Not far from IRS 9A,
\citet{2013MNRAS.433..712R} identified MTT 58 \citep{1989A&A...213...89M}
as an O2 star very close (in projection) to the molecular cloud core MM2E
(source \cc) seen first by \citet{2002A&A...394..253N} in C$^{18}$O (2--1)
data, indicating that this star, like IRS 9A, might still be embedded
in a parental gas and dust.  \citet{2013MNRAS.433..712R} estimated the
age of this star to be no more than 600 000 years, based on the size of
the associated \HII\ region. The age of the central star of IRS 9A
is about 70 000 years in the model 3012790 of \citet{robitaille2006} which
was fit by \citet{2010A&A...520A..78V} to the spectral energy distribution
of IRS 9A.


\section{Conclusions} \label{conclusions}

The SABOCA $350\mu$m image of the bright infrared source IRS 9A in the
nearby \HII\ region NGC 3603 confirms the presence of multiple massive
cores in the molecular clump MM2, while the SHFI spectra do not show the
typical hot core lines such as those from Methylcyanide.  Based on their
mid-infrared opacity, we classify these cores as infrared dark clouds
even though their temperature of about 50 K is higher than expected
for IRDCs.  The mid-IR source IRS 9A is associated with a massive core,
but outside it, and could be in a stage just after "hot core", i.e. an
ultra-compact \HII\ region, ionized by a massive star in its center.
Millimeter hydrogen recombination line emission was detected in
this direction, but also from a nearby core indicating a potentional
diffuse contribution due to the ionizing radiation from the cluster. It
is the only source in MM2 strong in the mid-infrared.  The structure of
MM2 as seen in the CS(2--1) line at $3$ mm by ATCA is characterized by
filaments at different systemic velocities.  The CO mapping data do not
show conclusive evidence for a high velocity outflow, certainly not for
a very massive one. However, the presence of SiO emission, with a hint
of line wing emission, is indicative of weak outflow activity.

\begin{acknowledgements}
This research has made use of the SIMBAD
database, operated at CDS, Strasbourg, France.
We thank the anonymous referee for comments which helped improve our paper.
\end{acknowledgements}


\end{document}